\documentstyle[12pt]{article}

\newcommand{\be}{\begin{eqnarray}}
\newcommand{\ee}{\end{eqnarray}}

\renewcommand{\vec}[1]{{\bf #1}}
\begin{document}
\begin{flushright}
Preprint ITEP--TH--21/96
\end{flushright}

\vspace{2cm}

\begin{center}
{\Large Two-Dimensional Instantons with Bosonization and
Physics of
Adjoint $QCD_2$.}
\vspace{1cm}

A.V. Smilga\\
\vspace{0.5cm}

{\it ITEP, B. Cheremushkinskaya 25, Moscow 117259, Russia}
\vspace{1cm}
\end{center}

\abstract{We evaluate partition functions $Z_I$ in
topologically nontrivial
(instanton) gauge sectors in the bosonized version of the
Schwinger model and
 in  a gauged WZNW model corresponding to $QCD_2$ with
adjoint fermions.
 We show that the
bosonized model is equivalent to the fermion model only if a
particular form
of the WZNW action with gauge-invariant integrand is chosen.
For the exact
correspondence, it is necessary to integrate over the ways
the gauge group $SU(N)/Z_N$ is embedded into the full
$O(N^2 - 1)$
group  for the bosonized matter field. For even $N$, one
should also
take into account  the
contributions of both disconnected components in $O(N^2 -
1)$.
In that case, $Z_I \propto m^{n_0}$ for small fermion
masses where $2n_0$  coincides with the number of fermion
zero modes in a
particular instanton background. The Taylor expansion of
$Z_I/m^{n_0}$
in mass involves only even powers of $m$ as it should.

The physics of adjoint $QCD_2$ is discussed. We argue that,
for odd $N$,
the discrete chiral symmetry $Z_2 \otimes Z_2$
present in the action is broken spontaneously down to $Z_2$
and the
fermion  condensate $<\bar{\lambda} \lambda>_0$ is formed.
The system
undergoes a first order phase transition at $T_c = 0$ so
that the condensate
is zero at an arbitrary small temperature.
It is not yet quite clear what happens for even $N \geq 4$.}

\section{Introduction.}
It is known for a long time that the Schwinger model
involves topologically
nontrivial gauge field configurations--- the instantons (see
\cite{inst} and
references therein). The reason why they appear is the
nontrivial
$\pi_1[U(1)] = Z$. Instantons are characterized by an
integer topological
charge
  \be  \label{nu}\nu\ =\  \frac 1{2\pi} \int d^2x \ F(x)\
\ee
  where $F = F_{01} = \partial_0 A_1 - \partial_1 A_0$.
Their physics is rather similar to the physics of instantons
in $QCD_4$
with one light quark flavor. In particular, the fermion
condensate
 \be  \label{condSM}
|<\bar {\psi} \psi>_0| = \frac
g{2\pi^{3/2}} e^\gamma  \ee
 is formed ($g$ is the coupling and $\gamma$ is the Euler
constant). The
 path integral calculation of $|<\bar {\psi}\psi>_0|$
\cite{indus,Wipf}
 follows closely the 't Hooft calculation of the instanton
determinant in
 $QCD_4$. The condensate is formed due to the presense of
one complex fermion
 zero mode for gauge field background configurations with
unit topological
 charge $\nu$. It was noted recently that topologically
nontrivial
 configurations appear also in non-abelian two-dimensional
gauge theories
 with adjoint matter content \cite{Witten,cond}. In this
paper, we will
consider only a simplest non-trivial theory of this kind
which involve
a multiplet of adjoint real fermions $\lambda^a$. The
lagrangian of the
model reads
\be
 \label{Lqcd2}{\cal L} = -\frac 1{4g^2} F_{\mu\nu}^a
F_{\mu\nu}^a +
 \frac i2\left\{{\lambda}^a_L[\delta^{ab}\partial_-
 -f^{abc}A_-^c]\lambda^b_L
+{\lambda}^a_R[\delta^{ab}\partial_+ -
 f^{abc}A_+^c]\lambda^b_R \right\}
 \ee
where $\partial_\pm = \partial_0 \pm \partial_1,\ \ A_\pm^c=
A_0^c \pm A_1^c$
 and $\lambda_{L,R} = \frac 12(1 \pm\gamma^5) \lambda$ are
the left moving
 and right moving components of the Majorana fermion field
(the lagrangian
 is written in Minkowski space because Majorana fermions
cannot be defined
 in Euclidean space \cite{Ramon}\footnote{Note, however,
that though we
 cannot define the Euclidean
counterpart of the lagrangian (\ref{Lqcd2}), the Euclidean
{\it path
integral} can be easily defined as an analytic continuation
of the
Minkowski path integral. In Minkowski space, integration
over Majorana
fermions provides the factor which is the square root of the
Dirac
determinant. We can {\it define} the Euclidean path integral
of the
theory (\ref{Lqcd2}) as the integral over gauge fields
involving the
square root of the Euclidean Dirac determinant as a factor
\cite{Vain}. The
extraction of square root presents no problem here as all
eigenvalues
of the Dirac operator for complex adjoint fermions are
doubly
degenerate \cite{cond,Leut}.}). We will consider both
massless model
(\ref{Lqcd2}) and the model which includes a small mass term
 \be
  \label{massf}
 m \bar \lambda^a \lambda^a \ = \ -2i  m\lambda^a_L
\lambda^a_R ,
  \ee
$m \ll g$.

 Adjointness of all fields in the lagrangian  is
 crucial for the instantons to appear: in the standard
$QCD_2$
 with fundamental quarks where the
gauge group is $SU(N)$, $\pi_1[SU(N)] = 0$ and
topologically nontrivial
configurations are absent. But in the theory with adjoint
matter the true
gauge  group is  $SU(N)/Z_N$ (the
elements of the center act trivially on adjoint fields).
$\pi_1[SU(N)/Z_N] = Z_N \neq 0$ and instantons appear.
\footnote{A nontrivial $\pi_1[SU(N)/Z_N]$ brings about
topologically nontrivial configirations also in 4-
dimensional Yang--Mills
theory without quarks. But here two extra transverse
dimensions are present
and these configurations are not localized and have infinite
action. These
planar instantons were obtained in Ref.\cite{Altes} and
misinterpreted as
real ``walls between different $Z_N$ phases''. Actually, the
instantons
and  planar instantons are essentially Euclidean
configurations and do not
exist as real physical objects in Minkowski space
\cite{bub}.} It was found
in \cite{cond} that these configurations involve fermion
zero modes (that conforms with the analysis by Kogan \cite{Kog} who showed that
instantons do not contribute in the partition function in the massless theory
(\ref{Lqcd2}) in high temperature region). For the
simplest topologically nontrivial sector their number is
$2(N - 1)$.
Instantons lead to physically observable effects (with an
obvious reservation
that we are discussing a model theory which is not found in
Nature).
They are responsible, in particular, for finite string
tension in fundamental
Wilson loop, i.e. for confinement of heavy fundamental
sources in a theory
with non-zero mass of dynamic adjoint fermions (in massless
theory,
instantons decouple, string tension disappears, and the
sources are not
confined but screened) \cite{conf}. When $N = 2$, instantons
bring about a
non-zero fermion condensate \cite{cond}.

The latter follows also from semi-heuristic arguments based
on bosonization
 approach. The bosonized version of $QCD_2$ with fermions in
the adjoint
 representation of $SU(2)$ is the gauged WZNW model
 \cite{WZ}--\cite{gaugeWZ} with the matter fields
presenting orthogonal
 matrices $h^{ab}(x)$, the elements of $ O(3)$. The theory
involves only
 massive excitations, their mass being of order of the
coupling
 constant $g$. As a result, the matter field is ``frozen''
and a non-zero
 vacuum expectation value $<h^{aa}>_0$ appears. In the
fermion language,
 that means the appearance of non-zero $<\bar{\lambda}^a
\lambda^a>_0$
where $\lambda^a$ are adjoint Majorana fermion fields.
For $N \geq 3$, the situation is much more complicated and
controversial.
Instantons involve ``too many'' fermion zero modes and
cannot generate a non-vanishing bilinear fermion condensate.
On the other hand,
the quoted bosonization arguments do not distinguish between
different $N$.
Say, for $N = 3$, the matter fields present $8 \times 8$
adjoint $SU(3)$
matrices and a non-zero
 \be  \label{condnab}
 <\bar{\lambda}^a \lambda^a>_0 \ = \ Cg <h^{aa}>_0
\ee
 should appear. It is also known that the condensate is
formed at infinite $N$ \cite{Kogan}.

This paradox formulated in \cite{cond} is akin to a similar
paradox which pops
out in 4D SUSY Yang--Mills theories with higher orthogonal
groups
\cite{SYMort}, is rather troublesome, and it is not yet
absolutely clear how
it is
resolved. It was the main motivation for the present study.

The main part of
the paper is devoted to the analysis of Euclidean path
integrals of the
gauged WZNW model  in the topologically nontrivial sectors.
We show that the
zero mode suppression factor $\propto m^{n_0}$ is reproduced
indeed,
but only if doing things with a proper care.

The commonly
used form
of the gauged WZNW action reads
  \be
  \label{actold}S_E[A, h] = \frac 1{4g^2} \int d^2x \
  F_{\mu\nu}^a F_{\mu\nu}^a \nonumber \\+ \frac 1{16\pi}
\int d^2x \ {\rm Tr}
  \{ \partial_\mu h\partial_\mu h^{-1} \} - \frac
i{24\pi}\int_Q d^3\xi
\ \epsilon^{ijk} \ {\rm Tr} \{h^{-1} \partial_i h\ h^{-1}
\partial_j h \ h^{-1}
  \partial_k h \} \nonumber \\
  + \frac 1{8\pi} \int d^2x \left[ {\rm Tr} \{A_+h\partial_-
h^{-1} \} +
  {\rm Tr} \{A_-h^{-1} \partial_+ h \}+ {\rm Tr} \{A_+ h A_-
h^{-1}\} -
  {\rm Tr} \{A_+ A_-\}\right] \nonumber \\
= \ \frac 1{2g^2} \int d^2x \ {\rm Tr} \ F_{\mu\nu}^2 +
\nonumber \\
N \left\{
 \frac 1{8\pi} \int d^2x \ {\rm Tr} \{ \partial_\mu u \
\partial_\mu u^{-1} \}  - \frac i{12\pi}\int_Q d^3\xi \
\epsilon^{ijk}
\ {\rm Tr} \{u^{-1} \partial_i u \
u^{-1} \partial_j u \ u^{-1} \partial_k u \}  \right.
\nonumber \\
\left. + \frac 1{4\pi} \int d^2x \left[ {\rm Tr} \{A_+u
\partial_- u^{-1} \} + {\rm Tr} \{A_-u^{-1} \partial_+ u \}+
{\rm Tr}
\{A_+ u A_- u^{-1}\} - {\rm Tr} \{A_+ A_-\} \right] \right\}
  \ee
where $h$ is the matrix $(N^2-1) \times (N^2-1)$ belonging
to the
adjoint representation of $SU(N)$ and $u$ is an associated
unitary matrix
$N \times N$:
 \be
 \label{hu}
h^{ab} = 2 {\rm Tr} \{t^a u t^b u^{-1}\},
 \ee
 $A_\mu$ are anti-hermitian matrices  $A_\mu = iA_\mu^a T^a$
and $T^a$  are the generators in a corresponding
representation,
$A_\pm = A_0 \pm
iA_1,\ \ \partial_\pm = \partial_0 \pm i\partial_1$, and $Q$
is a
three-dimensional manifold with a two-dimensional boundary
where the theory
actually lives. Our statement that,  generally speaking, the
action
(\ref{actold}) is {\it wrong}. It is not gauge--invariant
and does not
correspond to the original theory (\ref{Lqcd2}). One should
rather choose the
action  in the form
\be
\label{actnew}
S_E(F, u) =  \frac 1{2g^2} \int d^2x \ {\rm Tr} \
F_{\mu\nu}^2   \nonumber \\
-\frac N{8\pi} \int d^2x \ {\rm Tr} \{u^{-1} \nabla_\mu u\
u^{-1} \nabla_\mu u  \} - \frac {iN}{12\pi}\int_Q d^3\xi \
\epsilon^{ijk}
\ {\rm Tr} \{u^{-1}\nabla_i u
u^{-1} \nabla_j u \ u^{-1} \nabla_k u \}  \nonumber \\
+ \frac {iN}{8\pi} \int_Q d^3\xi \ \epsilon^{ijk}
\  {\rm Tr} \left \{ F_{ij} (u^{-1} \nabla_k u \ + \
\nabla_k u u^{-1} )
 \right\}
  \ee
  where
  $$ \nabla_i u = \partial_i u + [A_i, u], \ \ \ \\F_{ij} =
  \partial_i A_j - \partial_j A_i + [A_i, A_j] $$
  The functional (\ref{actnew}) was first written by Faddeev
\cite{Faddeev}.
  The actions (\ref{actold}) and (\ref{actnew}) differ by
the integral of a
   total derivative. For topologically trivial
configurations, this integral
    is zero and the actions (\ref{actold}) and
(\ref{actnew}) are equivalent,
but in the instanton sectors they are not. Actually, the
action (\ref{actold}) is not gauge-invariant in the
instanton sectors while
the explicit invariance of the {\it integrand} in
(\ref{actnew}) under the
gauge transformations
\be
\label{gaugetr}
A_\mu \ \to \ \Omega^{-1} (A_\mu + \partial_\mu) \Omega
\nonumber \\
u \ \to \ \Omega^{-1} u \Omega \nonumber \\ \ee
is seen immediately.
\footnote{The problem does not arise, of
course, in
$SU(N)$ WZNW models which are the most studied ones. They do
not involve instantons
and the action (\ref{actold}) is perfectly OK.}

Adding the mass term
\be
\label{massb}
\propto m \ {\rm Tr} \ h\  =\ 2m \ {\rm Tr}  \{u t^a u^{-1}
t^a\}
\ee
 in the action
(\ref{actnew}) and evaluating path integral, we will show
that the factor $m^{n_0}$ is singled out where $n_0$ is half
the number of
fermion zero modes
 \footnote{It is half the number not just the number because
we are dealing here with Majorana fermions and the fermion
path integral
provides the factor which is the square root of the
Euclidean Dirac
determinant. }

Unfortunately, it is not yet the end of the story. We will
see that the
action (\ref{actnew}) with the added mass term (\ref{massb})
{\it does}
not  exactly correspond to $QCD_2$ with massive adjoint
fermions.
Recall that the set of $N^2 -1$ of free adjoint fermion
fields is
habitually bosonized with the orthogonal matrices $h \in
O(N^2 -1)$
\cite{Witboson}.
For the theory involving gauge fields $\in SU(N)/Z_N$, one
should
rather use bosonization with adjoint $SU(N)$ matrices
$h^{ab} \in SU(N)/Z_N
\subset O(N^2 -1)$ \cite{Brown}. But there are many ways to
choose a subgroup $SU(N)/Z_N$
within the large orthogonal group. It turns out that, in
order to
preserve all symmetries of the fermion lagrangian and to get
a correct
mass dependence for the partition function {\it in
topologically
non-trivial sectors}, one has to average over all these
ways. In other
words, one has to write the mass term in the form
  \be
\label{massbR}
\propto m \ {\rm Tr} \ [h \in O(N^2-1)]  =\ 2m R^{ab}\ {\rm
Tr}
\{u t^b u^{-1} t^a\}
\ee
and average over all $R^{ab}$ belonging to the coset $O(N^2-
1)/[SU(N)/Z_N]$

The plan of the paper is the following. Before
proceeding with our analysis of bosonized theories, we
present in Sect.
2 a new derivation of zero mode counting rules in instanton
sectors in the
fermion language. Distinct topological sectors are labeled
by an integer
$k =0,1,\ldots, N-1$. In Ref. \cite{cond} only the case
$k=1$ (the instanton)
and $k = N-1$ (the antiinstanton) were analysed. For an
arbitrary $k$ the
result is
\be
  \label{n0}  n^0_L = n^0_R = k(N - k)  \ee
  Note that we are dealing here with an index theorem of new
  variety--- the number of left-handed and right-handed zero
  modes coincide and the conventional Atiah--Singer index
  vanishes.

In  Sect. 3 we start our analysis of bosonized
theories with a
  warm--up example of the Schwinger model. We will show that
correct results for the partition function in the instanton
sectors are reproduced indeed in bosonization language but
only if
choosing  the gauge-invariant form for the bosonized
lagrangian  depending
explicitly only on field strength $F$. We show that the
partition function in
the sector with topological charge $\nu$ involves a factor
$m^{|\nu|}$
reflecting the presense of $|\nu|$ zero modes in the fermion
description. \

In Sect. 4 we analyse gauged WZNW models with
the action
(\ref{actnew}). We show
that the contribution of the fields in the topological class
$k$ in the
partition function involves the factor $m^{k(N-k)}$ in
agreement with
the fermion counting
  (\ref{n0}). It also involves, however, the factor ${\cal
A}^{k(N-k)}$
where ${\cal A}$ is the total area of our manifold. That
implies the
constant asymptotics of the  correlator of $k(N-k)$ scalar
fermion currents at large
distances and
the existence of non-zero fermion condensate which seems to
be excluded by {\it
other} arguments.

In the first place, these are the arguments based on the
assumed
extensive form of the partition function $Z \propto
\exp\{-\epsilon_{vac} {\cal A}\}$ discussed earlier in
\cite{cond} and
anew in the end of sect. 4. Second, one can rigourously {\it
prove}
that the fermion condensate is absent in the high
temperature region
--- this is the subject of Sect. 5.

Possible ways to resolve the paradox are briefly discussed
at the end
of Sect. 4 and in more details --- in Sect. 6.
In particular,
an attractive possibility is that the fermion condensate
appears at $T=0$
due to spontaneous breaking
 of $Z_2 \otimes Z_2$ symmetry which the lagrangian
(\ref{Lqcd2}) enjoys: the
 transformations
 \be
 \label{Z2Z2}  \lambda_L \to - \lambda_L \nonumber \\
 \lambda_R \to - \lambda_R
\ee
 leave ${\cal L}$ invariant.
 This discrete  $Z_2 \otimes Z_2$ symmetry is the remnant of
$U(1)$ chiral
 symmetry which would be effective in a theory
 with complex fermions. A mass term (\ref{massf}) would
break this symmetry
 down to $Z_2$.
 And the appearance of the fermion condensate in massless
theory breaks
it spontaneously.

Spontaneous breaking of discrete symmetry would imply a
first order
phase transition at $T_c = 0$ (so that the condensate is
zero at any
non-zero temperature) --- much like in one--dimensional
Ising model. This picture is very much probable at $N=3$ and
at higher odd $N$, but the situation at even $N
\geq 4$ is not yet clear --- $Z_2 \otimes Z_2$ symmetry of
the lagrangian (\ref{Lqcd2}) is {\it anomalous} in this case
being broken explicitly by instanton effects.

In Sect. 7, we discuss the correspondence of the fermion and
the bosonized versions of the theory in more details. We
show that  the correct behavior of the fermion partition
functions in the instanton sector is reproduced only if
integrating the bosonized partition function over the
parameter $R \in O(N^2-1)/[SU(N)/Z_N]$ characterizing the
way the $SU(N)/Z_N$ subgroup is embedded in the larger
$O(N^2-1)$ group. This is the only way to enforce the
symmetry
(\ref{Z2Z2}) for odd $N$ in the bosonized version. For even
$N$, one has to take into account the contributions of both
disconnected components in $O(N^2-1)$.

Possible implications of our analysis for four-dimensional
supersymmetric gauge theories are discussed in the last
section.

\section{Index Theorem.}
\setcounter{equation}0
Instantons present a non-trivial fiber bundle $A_\mu(x)$ of
the gauge group
$SU(N)/Z_N$ on the 2-dimensional Euclidean manifold where
the theory is
defined. In Ref. \cite{cond} it was convenient to choose the
manifold to be a
torus. When the size of the torus in one of Euclidean
directions is small
compared to $g^{-1}$, the quasiclassical approximation works
and path
integrals in the instanton sector are saturated by fields at
the vicinity of a
 particular
configuration in the instanton class which has a very simple
abelian form.
In the case of large spatial volume and small temporal
size $\beta$
(that physically corresponds to high temperature $T=
\beta^{-1} \gg g$) the
relevant saddle point configuration in the topological class
$k=1$
is (the gauge $A_1 = 0$ is chosen)
 \be
  \label{Tinst}
  A_0(x) = \frac iN {\rm diag} (1,1,\ldots,1-N) \ a(x-x_0)
\ee
  where the profile function $a(x-x_0)$ has the same form as
in the
  Schwinger model \cite{inst} and the corresponding field
density
  $F =  - \partial A_0/\partial x$ is localized at the
vicinity of $x_0$,
   the instanton center. With the solution (\ref{Tinst}) at
hand, path
   integrals can be explicitly calculated and, for example,
the fermion
condensate in the high temperature limit can be found
\cite{cond,Thies}.
In \cite{cond} we explicitly solved the Dirac equation in
the background
(\ref{Tinst}) and found $N-1$ left--handed and $N-1$ right--
handed fermion
zero modes. We also showed that the eigenvalues do not shift
from zero when
perturbing the background (\ref{Tinst}) in every order of
perturbation
theory.
This reasoning was convincing enough but did not have the
rank of a rigourous
proof --- one could in principle contemplate the presence of
field
configurations in the instanton class at some distance in
Hilbert space from
the
abelian instanton (\ref{Tinst}) where the eigenvalue is
shifted from zero by non-perturbative effects. The main
problem here is that a standard Atiah--Singer index theorem
says nothing
about the presence or absence of these zero modes. The
Atiah-Singer index
is just zero here:
\be
\label{index} I^{\rm Atiah-Singer} = n_0^L - n_0^R
\sim {\rm Tr}\int F_{\mu\nu} \ \epsilon_{\mu\nu} \ d^2x = 0
\ee
A proof was constructed in \cite{Thies} where the theory was
studied on a finite spatial circle at zero temperature in
hamiltonian approach. In that case, the gauge $A_0 = 0$ can
be chosen and the dynamic variable is $A_1(x,t)$.
The point is that
the hamiltonian has $N$ classical vacua corresponding to
shifting $A_1$ from
zero by  particular finite constant matrices belonging to
Cartan subalgebra (see Sect. 5 for some more details).
The hamiltonian has a symmetry which guarantees that the
energy spectrum of
the Dirac operator in all classical vacua is identical. When
$A_1$
interpolates smoothly between adjacent vacua, exactly $N-1$
left-handed
levels with positive energy cross zero and go down into the
Dirac sea.
Likewise, $N-1$ right-handed levels from the sea cross zero
and appear
in the physical spectrum.
\footnote{Which levels --- left--handed or right--handed ---
go down into the
sea and which go out of it depends, of course,  on
convention and on the
direction in which $A_1$ is changed.} The level crossing
phenomenon
guarantees that the Euclidean Dirac operator has $N-1$
right--handed
and $N-1$ complex conjugated left--handed zero modes on any
background
which interpolates in Euclidean time between classical
vacua, i.e. on any
background belonging to the instanton topological class.

Both discussed proofs are somewhat indirect and we believe
it is worthwhile to
give a {\it direct} proof with explicit construction of the
zero mode solution.
Let us first derive the gauge field topological
classification more accurately. Topologically non-trivial
configurations exist only on compact Euclidean manifolds.
There are two
convenient choises --- a torus as in \cite{cond,Thies} or a
sphere. We
will return on torus in sect. 5, but currently we are moving
onto
sphere and will stay there for a while.
A sphere geometry appears when one  considers
 the gauge fields living on the Euclidean plane which tend
to a pure
 gauge at infinity:
 \be
 \label{asgauge}  A_\mu(x) \stackrel{r \to \infty}
{\longrightarrow}\
 \Omega^{-1}(\theta) \partial_\mu \Omega(\theta)  \ee
 with $\Omega(\theta) \in SU(N)/Z_N$. The matrix
$\Omega(\theta)$ defines a
 loop in the group space. Topologically non-trivial
configurations are
 described by
 non-contractible loops. The topological invariant
distinguishing different
 classes is
 \be
 \label{WC}W(C) = \frac 1N {\rm Tr} \exp \left\{
\oint_CA_\mu dx_\mu
 \right \} = \exp \left\{\frac{2\pi i k}{N}\right\}  \ee
 where the contour $C$ goes around the Euclidean infinity
and
 $k = 0, \ldots , N-1$. It is the same standard construction
as for the
 4-dimensional Yang--Mills instantons. The difference is
that in the latter
 case the topological invariant
 $$ I^{d =4} \sim \int_{S^3} K_\mu n_\mu $$
 can be written as a four--dimensional integral of the local
 topological charge density
 $\partial_\mu K_\mu \ \sim {\rm Tr}\{F_{\mu\nu}
\tilde{F}_{\mu\nu} \}$.
 On the other hand, the invariant (\ref{WC}) is inherently
non-local and
 cannot be presented as a two-dimensional integral of a
local density.
 Let us now choose a particular representative in each
topological class.
 A convenient choice is
 \be
 \label{A0k}A_\mu^{(0)k} = \frac iN {\rm diag}
 (\underbrace{k, \ldots,k}_{N-k}, \underbrace{k-N, \ldots ,
k-N}_k)
\ \frac {\epsilon_{\mu\nu}x_\nu }{(x_\mu^2 + \rho^2)}
  \ee
where we want to choose $\rho \sim g^{-1}$. This is a
configuration belonging to the class (\ref{WC}) with
localized field density
and finite action. For $k = 1$, the color structure of
(\ref{A0k}) is the
same as in (\ref{Tinst}).

We emphasize that (\ref{A0k}) {\it is} not a solution to the
classical equations of motion --- such a solution exists and
has the
 same color structure as (\ref{A0k}), but is delocalized:
the field density
 is constant on $S^2$ and very small, $F \sim 1/{\cal A}$
(${\cal A}$ is the
 area of the sphere).  Mathematically, this delocalized
configuration is
 as good a reference point as the configuration (\ref{A0k}).
The configuration
 (\ref{A0k}) is, however, preferable from the physical
viewpoint. Considering
  classical solutions makes sense only in the case when
quasiclassical
  description holds and characteristic fields in path
integrals are in
the vicinity of classical saddle points. However, $QCD_2$
at low temperature and large spatial volume  is a non-
trivial non-linear theory with
strong coupling
and the quasiclassical description is not adequate. An
analysis of the path
integral in the instanton sector shows that characteristic
field
configurations are actually localized at distances of order
of the
correlation length $\sim g^{-1}$ and resemble (\ref{A0k}) in
this
respect \cite{inst}.
\footnote{At high temperature $T \gg g$ quasiclassical
analysis becomes
possible which allows one to determine the value of the
fermion condensate for
$N=2$ \cite{cond,Thies}.
The saddle point field configuration of high $T$ path
integral in the
instanton sector presents the
solution of {\it effective} equations of motion with account
of the fermion
determinant. It has the form (\ref{Tinst}) and is localized
\cite{inst}.}

The field (\ref{A0k}) is defined on Euclidean plane and is
singular at
infinity. To define an instanton on the compact $S^2$
manifold, one should
either to use stereographic coordinates in which case the
field would be
singular at the north pole of the sphere or to go over in
the singular gauge
  \be
 \label{A0ksing}
A_\mu^{(0)k} = - \frac iN {\rm diag} (\underbrace{k,
\ldots,k}_{N-k},
\underbrace{k-N, \ldots , k-N}_k)\ \frac {\rho^2
\epsilon_{\mu\nu}x_\nu }{x_\mu^2(x_\mu^2 + \rho^2)} \ee
   (the size of the sphere $R$ is assumed to be much larger
than $\rho$).
The field (\ref{A0ksing}) has the same field strength $F$ as
(\ref{A0k}), is
regular at infinity and involves a Dirac string singularity
at $x = 0$.
Obviously, a gauge where the Dirac string is placed at any
other point $x_*$
on the sphere can be choosen.

Let us now solve the Dirac equation
  \be
  \label{Dirac}
\gamma_\mu^E \{ \partial_\mu \lambda_n \ +\ [A_\mu,
\lambda_n] \}\ =\ \mu_n \lambda_n
  \ee
with $\gamma_0^E = i\sigma^2,\ \ \gamma_1^E = i\sigma^1$,
$\mu_n$ being the eigenvalue corresponding to the eigenmode
$\lambda_n$,
on the background
(\ref{A0ksing}). Consider the
  matrix $\lambda^a t^a$. In Euclidean space,
Majorana fermions cannot be defined, and the fermion fields
should
be assumed to be complex. It is convenient to choose the
complex basis
$\{t^a\}$ for the Lee algebra with $N-1$ standard diagonal
matrices and
$N(N-1)/2 \ + \ N(N-1)/2$ off-diagonal matrices having only
one non-zero
component. In this basis,  the Dirac operator with abelian
background
(\ref{A0ksing})
does not mix the components $\lambda^a$ with different $a$
so that each
component can be treated separately. For some components,
the
commutator of the corresponding $t^a$ with the diagonal
color matrix in
(\ref{A0ksing}) is zero, these components do not feel a
background gauge
field at all, and the spectrum is the same as for free
fermions. An example
of the component which {\it does} feel the background is
\be
\label{tpstar}
(T_*^+)_{ij} \ = \left(
\begin{array}{c|c|c} i \setminus j &N-k & k\\
\hline
N-k& \mbox{\Huge O} &
\begin{array}{ccc}1 & ... &0\\ ... & ...& ... \\ 0 & ... & 0
\end{array} \\ \hline
k&\mbox{\Huge O} & \mbox{\Huge O
}\end{array} \right)  \ee
The Dirac equation for this component  looks the same as the
Dirac equation in
Schwinger model for the charged fermion field in the
background with unit
abelian topological charge (\ref{nu}). The standard Atiah--
Singer theorem
dictates the presence of a left-handed zero mode. Its
particular form is
\be
\label{ab0L}\lambda_{*L}^{(0)}(x)\  =\  T_*^+ {1 \choose
0}_{\rm
spin}
\frac {x_+}{\sqrt{x_\mu^2(x_\mu^2 + \rho^2)}}  \ee
There are $k(N-k)$ color matrices of the form (\ref{tpstar})
and,
correspondingly  $k(N-k)$ left-handed zero modes.
Also, there are  $k(N-k)$ right-handed zero modes
   \be
\label{ab0R}
\lambda_{*R}^{(0)}(x)\ = \ T_*^- {0 \choose
1}_{\rm spin}
   \frac {x_-}{\sqrt{x_\mu^2(x_\mu^2 + \rho^2)}}
  \ee
   where
     \be
     \label{tmstar}
     (T_*^-)_{ij} \ = \ \left( \begin{array}{c|c|c} i
\setminus j &N-k &
k\\
\hline N-k&\mbox{\Huge O} &
     \mbox{\Huge O} \\
\hline
k&\begin{array}{ccc}1 & ... &0\\
... & ...& ... \\
     0 & ... & 0 \end{array}  & \mbox{\Huge O }\end{array}
\right)
\ee
     etc. Up to now, we have just adapted the derivation of
\cite{cond} for
     the case when fields live on sphere and generalized
them on arbitrary $k$.
      In order to show explicitly the presence of
      $k(N-k) \ + \ k(N-k)$ zero modes on {\it any}
topologically nontrivivial
background, we use the fact that any field belonging to the
class $k$ can be
written as  \be
\label{decomp}
\left \{ \begin{array}{c} A_- = g^{-1} \left(\partial_-
+A_-^{(0)} \right) g\\
A_+ = g^{\dagger} \left(\partial_+ + A_+^{(0)}
\right)(g^\dagger)^{-1}
\end{array} \right.  \ee
where $g$ is a general complex $N \times N$ matrix. For a
unitary $g$ it is
just a gauge transformation. For a hermitian $g$ it is a
non-trivial
non-abelian field with a different field density but with
the same
invariant (\ref{WC}). We restrict, however, $g$ to be {\it
unitary} at
the point $x_*$ where the Dirac string is placed. To be
quite precise,
it is sufficient to require that $gg^\dagger$ commutes with
the
matrix marking out
the color direction of the Dirac string. Otherwise, the
transformed field
(\ref{decomp}) is not a fiber bundle on $S^2$

The decomposition (\ref{decomp}) is widely known for
topologically trivial
fields \cite{gaugeWZ}. It is a direct non-abelian analog of
the decomposition
 \be \label{decompab} A_\mu = A_\mu^{(0)} +
 \epsilon_{\mu\nu}\partial_\nu \phi+ \partial_\mu \chi  \ee
 of a topologically non-trivial field in the Schwinger model
on $S^2$
 \cite{indus}. Substituting (\ref{decomp}) in the Dirac
equation (\ref{Dirac}), one
can easily find the explicit expression for the zero modes
\be
\label{gen0}
(\lambda^{(0)}_L)_g = g^{-1} \lambda^{(0)}_L g \nonumber \\
(\lambda^{(0)}_R)_g = g^\dagger \lambda^{(0)}_R
(g^\dagger)^{ -1}  \ee
where $\lambda^{(0)}_{L,R}$ are the zero
modes(\ref{ab0L},\ref{ab0R}) for
the instanton representative (\ref{A0ksing})

\section{Instantons in Bosonized Schwinger Model.}
\setcounter{equation}0
Our main goal is to reproduce the zero mode counting of the
previous section
in bosonization approach. Of course, there is no trace of
fermion zero
modes in the bosonized theory. The proper question to ask is
how the
contribution to the partition function coming from instanton
sectors
depend on a small (smaller than any other relevant scale)
fermion mass $m$.
In the original theory with fermions, the behavior is
$Z_k \sim m^{k(N-k)}$. And the same should be true in the
bosonized WZNW model
--- the bosonized version of $QCD_2$. As a warm-up, consider
first the
abelian theory where the calculations can be carried out
explicitly until the
very end. The usual way to bosonize the Schwinger model is
to establish the
correspondence
\cite{Coleman}
\be
\label{corresp} i\bar\psi \partial_\mu \gamma_\mu \psi
\longrightarrow
\frac 12(\partial_\mu \phi)^2 \nonumber \\
 \bar\psi  \gamma_\mu \psi \longrightarrow \frac
1{\sqrt{\pi}}
 \epsilon_{\mu\nu} \partial_\nu \phi \nonumber \\
 \bar\psi  \psi \longrightarrow  -\frac
{e^\gamma}{2\pi^{3/2}}\ g\ \cos
 \left( \sqrt{4\pi}\ \phi \right)  \ee
 where $\gamma$ is the Euler constant. Then the Euclidean
action of
 the bosonized Schwinger model is
  \be
  \label{SMEwrong}  S_E = \int d^2x \left[ \frac 1{2g^2} F^2
+ \frac 12
  (\partial_\mu \phi)^2 + A_\mu \frac i{\sqrt{\pi}}
\epsilon_{\mu\nu}
  \partial_\nu \phi -  mg\ \frac {e^\gamma}{2\pi^{3/2}} \
\cos \left(
  \sqrt{4\pi}\ \phi \right)  \right]  \ee
  where $F = \epsilon_{\mu\nu} \partial_\mu A_\nu$ and
$\phi$ is a real scalar field .
Adding a full derivative to (\ref{SMEwrong}), one can
rewrite it in the form
  \be
  \label{SME}
  S_E = \int d^2x \left[ \frac 1{2g^2} F^2 + \frac 12
(\partial_\mu \phi)^2 +
  iF\phi\ \frac 1{\sqrt{\pi}}  -  mg\ \frac
{e^\gamma}{2\pi^{3/2}} \ \cos
  \left( \sqrt{4\pi}\ \phi \right)  \right]  \ee
  Our remark is that the transformation from
(\ref{SMEwrong}) to (\ref{SME})
  is innocent {\it only} in the topologically trivial gauge
sector.
  In instanton sectors, the integral of a full derivative
produces a surface
  term which contributes in the action and {\it can}not be
disregarded.
  To see
  that, it is convenient to think of an instanton  on $S^2$
as of a monopole.
  The flux (\ref{nu}) is then
  associated with the flux of the monopole magnetic field
through a
  sphere surrounding the magnetic charge in a fictitious
three-dimensional
  space, i.e. with the magnetic charge itself. The potential
$A_\mu(x)$ of our instanton/monopole should involve a
singularity (the
Dirac string) at some point $x_*$ on $S^2$ . The surface
term appears just
due to this Dirac string singularity and produces the term
$-i\sqrt{4\pi}\phi(x_*)$ in the action. Whenever this
matters,
it is the action
(\ref{SME}) which should be used, not (\ref{SMEwrong}).
Actually, the
action (\ref{SMEwrong}) is not gauge--invariant in
topologically nontrivial
sectors. The term $-i\sqrt{4\pi}\phi(x_*)$ by which
(\ref{SMEwrong}) differs
from the explicitly invariant action (\ref{SME}) depends on
the position of
the Dirac string singularity, i.e. on the gauge.
\footnote{Obviously, one can repeat this reasoning without
invoking the
Dirac string, but describing the instanton fiber bundle with
a couple
of maps which is more accurate from the mathematical
viewpoint. The
physical conclusion, however, is the same.}

A traditional way to handle the bosonized theory is to do
first the
Gaussian integral over $\prod dF$ to obtain
 \be
   \label{Sphi}
   S_\phi = \int d^2x \left[  \frac 12 (\partial_\mu \phi)^2
+
   \frac{g^2}{2\pi} \phi^2\   -  mg\ \frac
{e^\gamma}{2\pi^{3/2}} \ \cos
   \left( \sqrt{4\pi}\ \phi \right)  \right]  \ee
   It is OK as far as we are not interested in the
contribution of a
   particular gauge topological sector. In the latter case,
one should
   proceed more accurately. Let us consider the theory on a
compact
   two--dimensional Euclidean manifold with large but finite
area ${\cal A}$
   which we choose to be $S^2$. To single out the
contribution of a particular
   instanton sector, we impose the condition (\ref{nu}) The
topological
   charge $\nu$ is an integer. In the original fermion
theory, this follows
   from the necessity to define the Dirac operator on the
compact manifold in
   a background gauge field. The eigenfunctions and the
spectrum exist only
   for integer $\nu$. In the bosonized language,
quantization of $\nu$
   follows from an additional requirement that the action
(\ref{SME}) is
   invariant under the shift $\phi \to \phi+ \sqrt{\pi}$. $
\sqrt{\pi}$ is
   just the period of the cosine in Eq.(\ref{SME}). We will
shortly see that
   even if we would allow for
non-integer $\nu$'s, the contribution of such fields in the
partition function
 is zero.

 Let us now expand the fields $F(x)$ and $\phi(x)$ in the
series
 over spherical harmonics
   \be
   \label{harm}  F(x) = \sum_{lm} F_{lm} \ Y_{lm}(\theta,
\varphi) \nonumber \\
     \phi(x) = \sum_{lm} \phi_{lm} \ Y_{lm}(\theta, \varphi)
\ee
     The zero harmonic $F_0 = 2\pi\nu/{\cal A}$ is fixed due
to (\ref{nu}).
     Integrating out all other harmonics of the gauge field,
we obtain
       \be
       \label{Znu}
       Z_\nu \ =\  e^{- \frac{2\pi^2 \nu^2}{{\cal A}g^2}}
       \int_{-\sqrt{\pi}/2}^{\sqrt{\pi}/2} d\phi_0 e^{i\nu
\sqrt{4\pi}
\phi_0} \cdot \nonumber \\
       \int \prod d\tilde{\phi}(x) \exp \left\{ - \int_{S^2}
d^2x
       \left[ \frac 12 (\partial_\mu \tilde{\phi})^2 +
\frac{g^2}{2{\pi}}
       \tilde{\phi}^2\   -  mg\ \frac {e^\gamma}{2\pi^{3/2}}
\ \cos \left(
       \sqrt{4\pi}( \phi_0 + \tilde{\phi}) \right)  \right]
\right\}  \ee
  where $\phi_0$ is the zero harmonic of the matter field
and
  $\tilde{\phi}(x)$ is the sum of all the rest. The interval
of integration
  over $\phi_0$ is restricted due to the periodicity of the
integrand.
  It is instructive to see what happens if we sum over
$\nu$. Using the dual
  representation of
the $\Theta$--function, we obtain
\be
\label{Ztot}  Z = \sum_\nu Z_\nu \ \propto \
\int_{-\sqrt{\pi}/2}^{\sqrt{\pi}/2}
d\phi_0
\sum_{k =-\infty}^{\infty} \exp \left\{ - \frac {g^2 {\cal
A}}{2\pi}
(\phi_0 - k\sqrt{\pi})^2 \right \} \nonumber \\    \int
\prod d\tilde{\phi}(x) \exp \left\{ - \int_{S^2} d^2x \left[
\frac 12
(\partial_\mu \tilde{\phi})^2 + \frac{g^2}{2{\pi}}
\tilde{\phi}^2\
-  mg\
\frac {e^\gamma}{2\pi^{3/2}} \ \cos \left( \sqrt{4\pi}(
\phi_0 + \tilde{\phi})
\right)  \right]    \right\}
\ee
In the thermodynamic limit $g^2{\cal A} \to \infty$  only
one term of
the sum (\ref{Ztot}) survives, $\phi_0$ is frozen  at
zero, and
we reproduce the result (\ref{Sphi}). It is not difficult
also to calculate
the partition function in the theory with a particular non-
zero vacuum
angle $\theta$
\be \
\label{Ztet}
Z(\theta) = \sum_\nu Z_\nu e^{i\nu\theta}
\ee
Peforming the same dual transformation for this sum as for
$Z(0)$, we
arrive at the same expression  (\ref{Ztot}) but with the
shift
$\phi_0 \to \phi_0 + \frac \theta
{\sqrt{4\pi}}$. In that case $\phi_0$ freezes at the value
$\phi_0 = -\frac
\theta {\sqrt{4\pi}}$.

Now let us look at Eq. (\ref{Znu}). Note first of all that
though the bosonized
 {\it action} (\ref{SME}) is complex, the path integral
for $Z_\nu$ is real as
  it should. Second, we see immediately that in the massless
case $m = 0$,
  $Z_\nu = 0$ when $\nu \neq 0$. But for small but non-zero
$m$, $Z_\nu$ is
  non-zero too. A finite result is obtained when pulling
down the mass term in
   Eq.(\ref{Znu}) $\nu$ times. If we would try to calculate
$Z_\nu$ for a
   fractional $\nu$, the integral over $\phi_0$ would run
from $- \infty$ to
   $\infty$, the oscillating factor $\exp\{
i\nu\sqrt{4\pi}\phi_0\}$ could not
   be compensated in any order in $m$, and we would get zero
for any value of
   mass. This is the real reason for the topological charge
to be quantized:
   fractional topological charges just do not contribute
here in the partition
    function.
    \footnote{We hasten to comment that, in some theories
like twisted
multiflavor Schwinger model \cite{fracton} or four-
dimensional Yang--Mills
theory involving only adjoint color fields
\cite{toron,Leut},
fractional topological charges {\it do} contribute. In each
particular theory,
a particular study of this question is required.}
In the limit $mg{\cal A}\ll 1$ only the leading term in mass
expansion
survives (see \cite{Leut,inst} for a detailed discussion)
and we obtain
\be   \label{Znures}   Z_\nu = C_\nu (mg{\cal A})^\nu   \ee
with a calculable coefficient. This is exactly what we also
get in the
fermion language. For $\nu = \pm 1$, the coefficient $C_1 =
C_{-1} =
{e^\gamma}/(4\pi^{3/2})$ just gives the value of the fermion
condensate
(\ref{condSM}).

\section{Instantons in Gauged WZNW Model.}
\setcounter{equation}0
We have already mentioned that in topologically non-trivial
sectors it is
the action (\ref{actnew}) which should be used, not
(\ref{actold}).
The action (\ref{actnew}) relates to the action
(\ref{actold}) exactly
in the same way as the action (\ref{SME}) to
(\ref{SMEwrong}). The following
identity holds
 \be  \label{relat}
S_E^{{\rm Eq.(\ref{actnew})}}[A_\mu, u]\  =\
 \frac 1{2g^2}
 \int {\rm Tr}
\{F_{\mu\nu}^2\} -  \frac N{8\pi} \int d^2x \ {\rm Tr} \{
u^{-1}\nabla_\mu u\ u^{-1}\nabla_\mu u \} - \nonumber \\
-\ \frac {iN}{12\pi}\int_Q d^3\xi \ \epsilon^{ijk} \
{\rm Tr} \{u^{-1} \partial_i u \
u^{-1} \partial_j u \ u^{-1} \partial_k u \} \nonumber \\
+ \frac {iN}{4\pi} \int_Q d^3\xi \ \epsilon^{ijk} \partial_i
\ {\rm Tr}
\{ u A_j u^{-1} A_k +
A_j(u^{-1}\partial_k u + \partial_k u u^{-1}) \}
  \ee

Let us assume that the gauge fields have only two components
$A_0,\ A_1$
and depend only on the physical coordinates $x_\mu \equiv
\tau,\ x$.
The matter field $u(x_\mu, \alpha)$
is smooth on $Q$ and depends on the third coordinate $\alpha
\in [0,1]$ in
such a way that
$u(x_\mu,\ 0) = 1$ and $u(x_\mu,\ 1) $ is the field living
on our physical
2-dimensional Euclidean manifold ${\cal M}$ --- the boundary
of $Q$. One can
choose for example $u(x_\mu, \alpha) =
\exp\{\alpha \phi(x_\mu)\}$ with antihermitean $\phi$.

  For topologically trivial gauge fields which are regular
on ${\cal M}$, the integral of
  the full derivative is reduced to two surface terms at
$\alpha = 0$ and $\alpha = 1$ and produces together with
other terms the
standard form of the action
(\ref{actold}). But in instanton sectors,  fields involve
Dirac string
singularities on ${\cal M}$ which result in the additional
contribution in the
full derivative integral. For example, for $N=2$, the
relation
  \be
  \label{oldnew}
S^{Eq.(\ref{actold})} \ = \ S^{Eq.(\ref{actnew})} +  \ 2
{\rm Tr}
\{\phi(x_*)\  n^a t^a\}
 \ee
holds. Here $x_*$ is the position of the Dirac string and
$\vec{n}$ is its
direction in the color space. Obviously, the extra term in
(\ref{oldnew})
is gauge--dependent.

Let us now estimate the contribution of the instanton
sectors in the partition
function using the correct gauge-invariant expression
(\ref{actnew}) for the
action. An experience with Schwinger model teaches us that
the relevant factors
in the path integral appear due to integration over the zero
harmonic of the
matter field. Thus, we assume
  \be
  \label{uB}
u(x_\mu, \alpha) \ = \ \exp\{\alpha \beta\}
 \ee
where $\beta = i\beta^a t^a$ is a constant
antihermitean matrix.

Consider first the simplest case $N = 2$. The field has a
Dirac string
singularity at some point $x_*$ on $S^2$. We choose a gauge
with $x_* = 0$ and
direct the Dirac string along the third isotopic axis. The
singularity at small $x$ can be inferred from Eq.
(\ref{A0ksing}):
  \be
 \label{string2}
A_\mu^{sing}(x) \ = \ - it^3 \frac {\epsilon_{\mu\nu}
x_\nu}{x_\mu^2}
 \ee
A look at Eq.(\ref{actnew}) shows that the second and the
third terms in the
action may provide a divergent contribution $\propto \int
d^2x/x^2$ in the
action. Actually, the integral
$$ \propto \int_Q d^3\xi \ \epsilon^{ijk}  \ {\rm Tr} \{u^{-
1}\nabla_i u \
u^{-1} \nabla_j u \ u^{-1} \nabla_k u \}$$
{\it is} not divergent due to the fact
$\epsilon_{\mu\nu}A_\mu^{sing}
A_\nu^{sing} \ = \ 0$. But the integral
  \be
 \label{sing}
\propto \int d^2x \ {\rm Tr} \{u^{-1} \nabla_\mu u\
u^{-1} \nabla_\mu u  \} =
\int d^2x \ {\rm Tr} \{u^{-1} [A_\mu, u]\
u^{-1} [A_\mu u]  \}
  \ee
{\it is} singular provided $[A_\mu, u] \neq 0$. It would
give an infinite
contribution
in the action and the corresponding contribution in the
partition function is
suppressed. Thus, we should restrict ourselves with  the
constant
($x_\mu$ -- independent) matrices
(\ref{uB}) aligned in the same color direction as  the Dirac
string in a choosen
gauge. For such $u$, the only non-zero contribution in the
action comes from
the last term in (\ref{actnew}). We have
  \be
  \label{SB}
 S_E = -2 \ {\rm Tr} \{\beta t^3\} = -i\beta_3
  \ee
The instanton contribution in the partition function is
$$ Z_I \propto \int_0^{2\pi} d\beta_3 \exp\{-i\beta_3\} \ =
\ 0 $$
as it should be in the massless case [The range of $\beta_3$
is
restricted to be $[0,2\pi]$ because changing $\beta_3$ from
0 to $2\pi$
multiplies $u$ by the element of the center $-1$, and we
arrive at the
same associated orthogonal matrix (\ref{hu})]. If the
fermion mass is not zero,
the action involves an additional term
\be
\label{Sm}
 S_m \propto mg \int d^2x\ {\rm Tr}\ h(x) \ \propto mg{\cal
A}
[|{\rm Tr}\ u|^2 -1] = mg{\cal A} (2\cos \beta_3 + 1)
                                                   \ee
where ${\cal A}$ is the area of the manifold. Pulling the
mass term down, we
get in the leading order in $m$
 \be
 \label{ZI2}
Z_I
\ \propto \ mg{\cal A}
\int_0^{2\pi} d\beta_3 \exp\{-i\beta_3\}\ (2 \cos \beta_3 +
1) = Cmg{\cal A}
  \ee
with a nonzero constant $C$. That agrees well with the
results of the analysis
in the fermion language: a couple of fermion zero modes
provide a factor
$\propto m$ in the partition function. Differentiating
(\ref{ZI2}) over mass
gives the fermion condensate \cite{cond}.

Consider now the general color group $SU(N)$ and the field
configuration
of the type (\ref{A0ksing}) belonging to the topological
class $k$.
For any configuration in this class a gauge can be chosen
where the Dirac
string is aligned in the direction
 \be
  \label{Tstar}
T^* = \frac 1{\sqrt{2Nk(N-k)}} \ {\rm diag} (\underbrace{k,
\ldots,k}_{N-k},
\underbrace{k-N, \ldots , k-N}_k)
  \ee
in the color space. As earlier, we must require that the
constant mode of the
matter field $u_0$ commutes with $T^*$ --- otherwise the
second term in
(\ref{actnew}) would give an infinite contribution in the
action. A general
$u_0(\alpha = 1)$ satisfying this restriction has the form
  \be
  \label{u0k}
  u_0(1) \ = \ \exp\{i\beta^* T^*\}
\left( \begin{array}{cc} u^{(N-k)} & 0 \\ 0 & u^{(k)}
\end{array}
\right)
       \ee
where $u^{(N-k)} \in SU(N-k)$ and $u^{(k)} \in SU(k)$. We
assume $u_0(\alpha) = [u_0(1)]^\alpha$ so that $u_0(0) = 1$.
The parameter $\beta^*$
varies within the limits $\beta^* \in [0,\ 2\pi\sqrt{2k(N-
k)/N}]$ --- the shift
of $\beta^*$ by $2\pi\sqrt{2k(N-k)/N}$ multiplies $u_0(1)$
by an  element of the
center $\exp\{2\pi ik/N\}$ which results in the same adjoint
matrix $h$.
In the massless case, the only contribution in the action
comes from the last
term in (\ref{actnew}). It does not depend on $u^{(k)}$ and
$u^{(N-k)}$, but
only on $\beta^*$ and we have
$$
 Z_I^k \ \propto \ \int_0^{2\pi\sqrt{2k(N-k)/N}} d\beta^*
\exp\left\{-i\beta^* \sqrt{\frac {Nk(N-k)}2} \right\} \ = \
0
 $$
Note that the phase factor winds by $2\pi$ \ $k(N-k)$ times
in the range
of the integration.

The action in the massive theory involves the term
  \be
\label{Smk}
 S_m \propto mg \int d^2x\ {\rm Tr}\ h(x) \ \propto mg{\cal
A}
[|{\rm Tr}\ u_0|^2 -1] = mg{\cal A}
\left[
|{\rm Tr}\ u^{(k)}|^2 \right. \nonumber \\
\left. + |{\rm Tr}\ u^{(N-k)}|^2 +
2{\rm Re} \left({\rm Tr}\ u^{(k)} \ ({\rm Tr}\ u^{(N-k)})^*
\ \exp\left\{ -i\beta^*
\sqrt{\frac {N}{2k(N-k)}} \right\} \right) \ -\ 1\right]
 \ee
To provide a nonzero contribution in the path integral for
$Z_I^k$,
the mass term should be pulled
down at least $k(N-k)$ times --- otherwise the integral over
$\beta^*$ gives
zero. Note that not only $\int d\beta^*$ but also group
integrals over
$u^{(k)}$ and $u^{(N-k)}$ provide here non-zero factors.
Thus, we get an estimate
  \be
 \label{ZIk}
Z_I^k \ \sim (mg{\cal A})^{k(N-k)} \int_0^{2\pi \sqrt{2k(N-
k)/N}}
\int du^{(k)} ({\rm Tr}\ u^{(k) *})^{k(N-k)} \cdot \nonumber
\\
\int du^{(N-k)} ({\rm Tr}\ u^{(N-k)})^{k(N-k)}
 \sim (mg{\cal A})^{k(N-k)}
 \ee
for small $mg{\cal A}$.

The factor $m^{k(N-k)}$ appears also in the fermion approach
--- $k(N-k)$ is
just the number of the fermion zero mode pairs.
\footnote{As has already been mentioned in the Introduction,
the
bosonized theory with the action (\ref{actnew}) still does
not exactly
correspond to the original fermion theory. It is convenient
for us to
postpone the discussion of this issue till Sect. 7.}
 What is, however, new and
could not be figured out in the fermion approach is the
total area dependence
$\propto {\cal A}^{k(N-k)}$. Consider e.g. the case $N = 3$.
The instanton
partition function can be written as
 \be
 \label{ZI3}
Z_I^{N=3} \ \sim \ (mg)^2 \int d^2x \ d^2y \ <\bar \lambda^a
\lambda^a (x)\
\bar \lambda^a \lambda^a (y)> \ \sim \ (mg{\cal A})^2
  \ee
The appearance of the factor ${\cal A}^2$ in this expression
means that
the correlator \\ $<\bar \lambda^a \lambda^a (x)\
\bar \lambda^a \lambda^a (y)>$ tends to a nonzero constant
at large Euclidean
distances $|x-y|$, i.e. that the fermion condensate
$<\bar \lambda^a \lambda^a>$
is formed.

Thus, a bosonization estimate for $Z_I^k$ presented in this
section has
confirmed the existence of $k(N-k) \ + \ k(N-k)$ fermion
zero modes in
the path
integral and, on the other hand, confirmed the appearance of
the fermion
condensate for any $N$ which also follows from simplistic
bosonization
arguments of
Ref. \cite{cond}. This is rather remarkable, but
unfortunately  does not mean
yet that the physical
situation is now absolutely clear and a final resolution of
the paradox
mentioned in \cite{cond}
[the conflicting results of the bosonized
analysis and the fermion analysis of the theory
(\ref{Lqcd2}) for higher gauge
groups] is achieved.

The paradox displays itself if recalling the fact that the
spectrum of the
theory (\ref{Lqcd2}) does not involve massless particles.
That means that in
the limit ${\cal A}g^2 \gg 1$ when the size of the Euclidean
box is much larger
than the characteristic mass scale $\sim g$, the partition
function must enjoy
the extensive property
 \be
 \label{extens}
Z \ \propto \ \exp\{-\epsilon_{vac}(m,g){\cal A} \}
  \ee
and the finite volume corrections ( the boundary effects)
should be
exponentially suppressed \cite{Lee}. At small $m \ll g$,
$\epsilon_{vac}(m,g)$
should involve the linear in mass term --- the corresponding
coefficient just
gives the fermion condensate $=\  -1/{\cal A}\
\partial/\partial m\ \ln\ Z$
the existence of which is dictated by the estimates
(\ref{ZIk}), (\ref{ZI3})
for the instanton contribution in the partition function.
\footnote{For a related discussion in  $QCD_4$ see
\cite{Leut}.}

The property (\ref{extens}) should hold both in the true
thermodynamic limit
$mg{\cal A} \gg 1$ and also in the region $mg{\cal A} \ll 1$
provided the
condition ${\cal A}g^2 \gg 1$ is fulfilled. But, on the
other hand, for $N \geq 3$,
no known contribution in the partition function involves the
linear term $\propto
mg{\cal A}$ and the expansion of $Z$ in small $mg{\cal A}$
starts with the
 term $\sim m^{N-1}$.

There are only two ways out of this obvious contradiction:
 \begin{enumerate}
\item Perhaps, for some reason, topological classification
does not hold in this
case, and, besides instantons, there are some {\it other}
contributions in the
partition function which involve a linear in mass term and
would be responsible
for the formation of the fermion condensate in the limit
$mg{\cal A} \ll 1$.
These non-descript contributions would play the same role as
the toron (or meron
or fracton) contributions which are responsible for the
formation of the gluino
condensate in $SU(N)$ supersymmetric 4D Yang--Mills theory
\cite{toron} and the formation of the fermion condensate in
multiflavor Schwinger model in finite volume with twisted
boundary conditions \cite{fracton}.
This is the possibility advocated for in \cite{cond}.
\item Another possibility is that the topological
classification is good, the
``fracton'' contributions are absent and the partition
function {\it does} not
have an extensive form (\ref{extens}) for small $mg{\cal
A}$. But that
necessarily implies the existence of massless states in the
spectrum. As there
are no massless {\it particles}, the only choice is that the
vacuum state
involves a discrete degeneracy which is lifted by a small
fermion mass. Then
the {\it physical} partition function presents the sum of
two extensive
exponentials
 \be
 \label{Zsum}
 Z \ \sim \ \exp\{-[\epsilon_0 - Cmg + O(m^2)]{\cal A}\} \ +
\exp\{-[\epsilon_0 + Cmg + O(m^2)]{\cal A})\} \
 \ee
and the linear in mass term cancels out.
 \end{enumerate}

At present, we do not know what the answer is. We will
discuss these two
options in details in Sect. 6 and in the last section. But
before that, let
us discuss the
physics of the theory (\ref{Lqcd2}) at finite temperatures
where {\it definite}
conclusions can be done.

\section{Adjoint $QCD_2$ at High Temperature.}
\setcounter{equation}0

The main subject of this paper is analyzing the dynamics of
adjoint $QCD_2$
in bosonization approach. However, it is difficult to do at
finite temperature.
The reason is that, in contrast to $S^2$, a torus where a
finite
temperature theory is defined presents  not a simply
connected
manifold,
there are no smooth 3-dimensional manifolds parametrized by
a parameter
$\alpha \in [0,1]$ such that the value $\alpha = 0$
corresponds to a
single point on the manifold and the value $\alpha =1$
corresponds to
 the boundary which is torus. That brings about
problems with defining the Wess--Zumino term \cite{WitNWZ}.
Thus, we have to use the original fermion language.

The dynamics of the theory (\ref{Lqcd2}) at high temperature
$T \gg g$
for $N = 2,3$ was discussed at length in \cite{cond}. In
\cite{Thies}
the same theory was studied at $T = 0$ but on a small
spatial circle
$L \ll g^{-1}$ in hamiltonian approach. In Euclidean
approach the first
theory is defined on a cylinder $0 \leq \tau \leq \beta =
T^{-1},\
-\infty < x < \infty$ (for the theory to be completely
regularized in
the infrared, one may restrict also the range of $x$: $-L
\leq x \leq L$, but
the length of the box $L$ should be assumed to be very large
--- larger than
any relevant physical parameter), while the second theory is
defined on
a cylinder $-\infty < \tau < \infty,\ 0 \leq x \leq L$.
Obviously, the
both cases are completely equivalent up to interchange $x
\leftrightarrow \tau$.

Let us briefly summarize the results of these studies. We
will use
mainly the hamiltonian finite spatial circle language which
is a little
more transparent physically. Eventually, however, we are
going to
translate the results obtained in the finite temperature
language.

Consider first the simplest case $N=2$. Choose the gauge
$A_0 = 0$. The
dynamic variables are $A_1(x)$. In finite spatial volume,
the  zero
Fourier mode $A_1^{(0)}$ of the field $A_1(x)$ plays a
crucial role.
Actually, in the limit $gL \ll 1$, all other components and
the fermion
fields present the ``fast variables'' in the Born--
Oppenheimer approach
which have high characteristic excitation energies and can
be
integrated out. We are left with the effective potential
$V^{eff}(A_1^{(0)})$
depending on the slow variable $A_1^{(0)}$. $V^{eff}$ does
not depend on
isotopic orientation of $A_1^{(0)}$. For definiteness, we
may direct
it along the third isotopic axis: $A_1^{(0)} = i A_1^3 t^3$.
The effective potential has the form \cite{Kut,Kog}
  \be
    \label{Veff2}
    V^{eff}(A_1^3)\  = \ \frac {L}{2\pi} \left[ \left(A_1^3
+ \frac \pi{L}
    \right)_{mod. 2\pi/L} - \frac \pi{L} \right]^2
    \ee
It is periodic in $A_1^3$ with the period $2\pi/L$ and has
minima at
$A_1^3 = 2\pi n/L$ with integer $n$. The points $A_1^3 = 0$
and $A_1^3 =
2\pi/L$ can
be related by a gauge transformation:
  \be
    \label{gauge2}
i    \frac {2\pi}{L}     t^3 \ = \ \Omega^\dagger(x)
\partial_x \Omega(x),
    \ \ \ \ \Omega(x) \ = \ \exp\left\{ \frac{2\pi
ix}{L}t^3\right\}
    \ee
The unitary matrix $\Omega(x)$ is changed    from $\Omega(0)
= 1$ to
$\Omega(L) = -1$. The associated adjoint matrix $\in SO(3)$
[recall
that for the theory involving only adjoint fields the true
gauge group
is $SU(2)/Z_2$ rather than just $SU(2)$] makes a closed loop
in the group
which cannot be contracted to zero. Thus, (\ref{gauge2}) is
a
{\it large}
gauge transformation which cannot be continuously deformed
to zero, and
the point $A_1^3 = 2\pi/L$ presents a topologically non-
trivial
classical vacuum. Note that the configuration $A_1^3 =
4\pi/L$
corresponds to a gauge transformation $\Omega(x) =
\exp\{4\pi
ixt^3/L\}$ which can be continuously deformed to zero and is
a {\it
trivial} gauge copy of $A_1^3 = 0$.

The physical picture is very much similar to the vacuum
structure in $QCD_4$
\cite{Callan}. The only difference is that here we have not
infinitely
many but just 2 topologically distinct vacua. An Euclidean
field
configuration which interpolates smoothly between $A_1^3 =
0$ at $\tau
= -\infty$ to $A_1^3 = 2\pi/L$ at $\tau = \infty$ presents
an
instanton we were talking about before. It has 1 left--
handed and 1
right--handed fermion zero mode which give rise to a non-
vanishing
fermion condensate. An accurate calculation
\cite{cond,Thies} gives
  \be
   \label{condL}
  |< \bar\lambda^a \lambda^a>| \ = \ \frac
{8\pi^{3/2}}{gL^2}
\exp \left\{ -
\frac{\pi^{3/2}}{gL} \right\}
   \ee
 This explicit formula is valid in the region $gL \ll 1$
when the
Euclidean tunneling trajectory in the potential
(\ref{Veff2})
 has large action $\frac{\pi^{3/2}}{gL}$ and
 the quasiclassical approximation works. But a non--
vanishing fermion
condensate exists at any $L$ (at any temperature). At $L =
\infty$ ($T =
0$) it is estimated to be of order $g$. The condensate
depends smoothly
on $L$ (on $T$) and there is no phase transition.

The large gauge transformation presents an extra discrete
symmetry of the hamiltonian. Like in $QCD_4$, the proper way
of handling the theory is to impose a superselection rule
and divide the Hilbert space of the systems in two sectors
involving the states which are symmetric under such a
transformation and the states which are antisymmetric. The
partition functions in these sectors are
 \be
 \label{Zpm}
Z_+ \ =\ Z_{triv} \ +\ Z_I \nonumber \\
Z_- \ =\ Z_{triv} \ -\ Z_I
\ee
This is quite analogous to choosing a particular value of
$\theta$ in $QCD_4$, only in this case with only two
classical vacuum states the parameter $\theta$ can acquire
only two disctrete values: $\theta = 0$ and $\theta = \pi$.
The fermion condensate has opposite sign in these two
sectors.
Let us turn now to the simplest paradoxical theory with
$N=3$. Again,
in the limit $gL \ll 1$, the low energy dynamics of the
theory can be
described by the effective potential $V^{eff}(A_1^{(0)})$.
The constant
mode $A_1^{(0)}$ can be chosen to be a diagonal matrix
 \be
 \label{diag}
 A_1 \ =\  i\ {\rm diag} (a_1, a_2, a_3)\ \ \ \ \ \ \ \sum_i
a_i =
0
 \ee
 The potential has the form \cite{Kut,Kog}
 \be
 \label{Veff3}
 V^{eff}(a_i)\ =\ \frac {L}{2\pi} \sum_{i>j}^3
 \left[ \left(a_i - a_j + \frac \pi{L}
    \right)_{mod. 2\pi/L} - \frac \pi{L} \right]^2
\ee

\unitlength=0.6mm

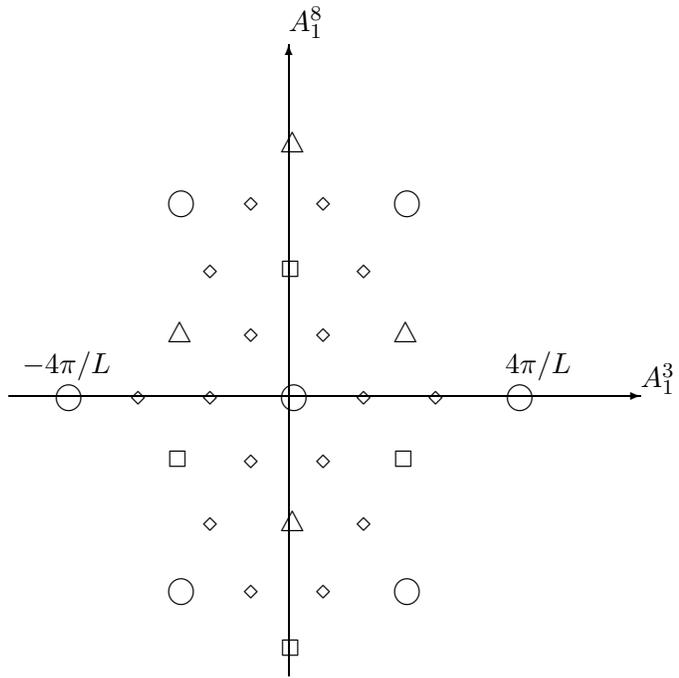
\begin{figure}
\begin{center}
\begin{picture}(160,160)
\put(10,72){\vector(1,0){140}}
\put(150,74){$A_1^3$}
\put(72,10){\vector(0,1){140}}
\put(72,153){$A_1^8$}

\put(70,70){$\bigcirc$}
\put(120,70){$\bigcirc$}
\put(20,70){$\bigcirc$}

\put(95,113){$\bigcirc$}
\put(95,27){$\bigcirc$}

\put(45,113){$\bigcirc$}
\put(45,27){$\bigcirc$}

\put(95,84){$\triangle$}
\put(45,84){$\triangle$}

\put(70,42){$\triangle$}
\put(70,126){$\triangle$}

\put(95,56){$\Box$}
\put(45,56){$\Box$}

\put(70,98){$\Box$}
\put(70,14){$\Box$}

\put(87,70){$\diamond$}
\put(103,70){$\diamond$}
\put(53,70){$\diamond$}
\put(37,70){$\diamond$}

\put(87,98){$\diamond$}
\put(53,98){$\diamond$}
\put(87,42){$\diamond$}
\put(53,42){$\diamond$}

\put(78,84){$\diamond$}
\put(78,56){$\diamond$}
\put(62,84){$\diamond$}
\put(62,56){$\diamond$}
\put(78,113){$\diamond$}
\put(62,113){$\diamond$}
\put(78,27){$\diamond$}
\put(62,27){$\diamond$}

\put(120,77){$4\pi/L$}
\put(13,77){$-4\pi/L$}

\end{picture}

\end{center}
\caption{Classical vacua in $N=3$ theory.}
\end{figure}
The pattern of its minima is shown in Fig. 1.
First, there are global minima divided in three topological
classes
(they are marked out by circles, boxes and triangles in
Fig. 1).
Each circle is  gauge
equivalent to any other circle with a
topologically trivial gauge transformation. The same is true
for boxes and  triangles.  The minima of
different
types are also gauge equivalent \, but with a topologically
nontrivial {\it large} gauge transformation.
An Euclidean
field configuration interpolating, say from $A_1 = 0$ at
$\tau =
-\infty$ to $A_1 = \frac {2\pi i}{3L} \ {\rm diag} (1, 1, -
2)$
at
$\tau = \infty$ presents an instanton. It has
2 left-handed and 2 right-handed zero modes which is too
much for the
fermion condensate to be formed. Like in the case $N=2$ and
like in $QCD_4$, the Hilbert state of the system can be
separated now in {\it three} sectors:
 \be
  \label{Zthet3}
 Z(\theta) \ =\ \sum_{k = 0,1,2}\ Z_k\ \exp\{ik\theta\}
 \ee
where $\theta = 2\pi n/3$  and $n = 0,1,2$. (Generally,
there are $N$ sectors with $\theta_n \ =\ 2\pi n/N$,\  $n =
0,1,\ldots,N-1$.

It was observed in
\cite{Thies} that,
besides global minima, the potential (\ref{Veff3}) has also
{\it local} minima
marked out with diamonds in Fig.1. The value of the
potential at
diamond points is
 \be
  \label{local}
  V_\diamond \ = \ \frac{2\pi}{3L}
  \ee
 We will shortly see that this value has some physical
relevance.

 The conclusions about the fermion condensate (its presence
at $N=2$
and its absence at $N \geq 3$) can be also reached in a
slightly different
way of reasoning which does not invoke instantons at all.
Consider the
correlator
 \be
 \label{corrt}
C(\tau) \ =\ <\bar \lambda^a \lambda^a (\tau) \ \bar
\lambda^a \lambda^a (0)>_L
 \ee
at very large $\tau$. In the limit of large $g^2{\cal A}$
and for $N \neq 3$, it is given by the path integral where
gauge
fields are {\it topologically trivial} (see \cite{indus} for
a detailed related discussion in the Schwinger model).
For $N = 3$, also instanton sectors contribute in the
correlator. But, as we will shortly see, the instanton
contributions can be analyzed along the same lines as the
topologically trivial contribution, and its behavior is also
the same. Consider first the
case $N=2$.
At small $gL$ the quasiclassical approximation is valid and
the correlator
 is mainly determined by the saddle point of the path
integral. This saddle
 point presents an {\it abelian} configuration
 \be
 \label{ab}
 A_1(\tau') \ =\ if(\tau') t^3
  \ee
 (It can of course be also rotated by a global gauge
transformation,
and it is important to take into account in a precise
calculation,
but for our purposes it is irrelevant.
The prime superscript is put to distinguish the running
argument $\tau'$ of
the profile function from the point $\tau$ where the second
fermion scalar
current is defined and on which the correlator (\ref{corrt})
depends).
The calculation of the correlator
(\ref{corrt}) on abelian background (\ref{ab}) is a simple
problem. The
point is that the component $\lambda^3$ does not feel the
background
and the term \\
$<\bar \lambda^3 \lambda^3 (\tau) \ \bar \lambda^3 \lambda^3
(0)>$
in the correlator (\ref{corrt}) is just the free fermion
correlator. In
finite box it decays  exponentially at large $\tau$:
\footnote{If going over in the finite temperature
interpretation,
$\tau$ is substituted by x, and the
factor $2\pi/L \equiv 2\pi T$ in the exponent is just twice
the lowest fermion
Matsubara frequency.}
  \be
  \label{free}
  C_{free}(\tau) \propto \exp\left\{- \frac {2\pi \tau}{L}
\right\}
  \ee
  and is irrelevant at large $\tau$.
The components $\lambda^1$ and $\lambda^2$ behave as a real
and
imaginary part of the Dirac fermion field having the abelian
charge
$g$ in an abelian gauge field background
$A_1(\tau') = f(\tau')$. Thus, the problem is reduced to the
abelian
Schwinger model problem. The behavior of the fermion
correlator  in
Schwinger model on the circle is well known . At large
$\tau$,
it tends to a constant. By
cluster decomposition, one can infer from this that a
fermion
condensate is formed both in the Schwinger model
\footnote{It is exactly the way
the expression (\ref{condSM}) for the fermion condensate in
the Schwinger model was originally derived \cite{condSM}.}
and in adjoint $QCD_2$.

Bearing in mind the generalizations which follow shortly,
let us give a brief
sketch how the result about the constant asymptotics of the
fermion
correlator in the Schwinger model is obtained (see e.g.
\cite{indus,Wipf,inst})
for more details). Use the decomposition (\ref{decompab}).
In the
topologically trivial sector, $A^{(0)}_\mu = 0$. The
gauge-dependent
part $\partial_\mu \chi$ is also irrelevant. The field
$\phi(x)$ is a
non--trivial gauge--independent degree of freedom and is
called {\it
prepotential}. In two dimensions, there
exists an exact formula for the fermion Green's function in
an arbitrary
background $\phi(x)$:
 \be
  \label{Green}
  S_\phi(x, y) \ =\ <\bar \psi(x) \psi(y)> = \exp\{-
g\gamma^5
\phi(x)\}
S_0(x-y) \exp\{-g\gamma^5 \phi(y)\}
 \ee
 where $S_0(x-y)$ is the free fermion Green's function.
Using
(\ref{Green}), we get
 \be
 \label{corrSM}
C^{SM}(x) \ =\ <\bar \psi \psi(x) \ \bar \psi \psi(0)>\
\propto \nonumber \\
C_{free}(x)
\prod d\phi \exp \left\{ - \frac 12 \int \phi(\Delta^2 -
\mu^2 \Delta)
\phi \ d^2y \right\} \cosh\{2g[\phi(x) - \phi(0)]\}
 \ee
 where $\mu^2 = g^2/\pi$ is the mass of the physical scalar
particle in
the spectrum (which may also be called heavy photon).
Performing the Gaussian
integration over $\prod d\phi(x)$, we obtain
for the correlator at large Euclidean time $\tau$ in the
theory defined
on a cylinder with small spatial size $L$
  \be
\label{corrtSM}
C^{SM}(\tau) \ =\  C_{free}(\tau) \exp\{4g^2 [{\cal G}(0)
- {\cal G}(\tau)]\}
  \ee
  where ${\cal G}(x)$ is the Green's function of the
operator
  $\Delta^2 - \mu^2 \Delta$ on a cylinder. The free
correlator falls
down exponentially at large $\tau$ according to (\ref{free})
while the second
factor rises
 \be
  \label{rise}
  \exp\{2g^2 [{\cal G}(0) - {\cal G}(\tau)]\} \ \propto \
\exp \left\{
\frac{2g^2}{\mu^2} \tau/L \right\} \ = \ \exp\{2\pi \tau/L\}
\ee
We see that the exponential decay of the free correlator is
exactly compensated
by the rising factor (\ref{rise}) and the correlator tends
to a constant
at large $\tau$.

Consider now the correlator (\ref{corrt}) in the theory with
$N=3$.
Again, for small $gL$ the quasiclassical approximation works
and the path
integral for the correlator is saturated by its saddle point
which is
abelian. A global $SU(3)$ rotation brings the potential
$A_1(\tau)$ in
a diagonal--color--matrix form. Saddle points appear in
different color
directions which are actually just the symmetry axes of
the effective
potential (\ref{Veff3}) and can be easily inferred from Fig.
1.
 Two essentially different options are $A_1^{saddle}(\tau')
= if(\tau') t^3$
 and $A_1^{saddle}(\tau') = ig(\tau') t^8$.

 Consider first the second case.
 If the gauge field is directed along the 8-th color axis,
the fermion
components $\lambda^{1,2,3,8}$ do not feel the field at all
and the
corresponding correlator has the asymptotics (\ref{free})
and is
suppressed compared to the contribution of other components.
The
components $\lambda^{4\pm i5}$ and $\lambda^{6\pm i7}$
interact with the background $A^8(\tau)$
as two complex fermions  of charge $g\sqrt{3}/2$ with an
abelian gauge
field background. Thus, the correlator $C(\tau)$ behaves at
large
$\tau$ exactly in the same way as the fermion correlator in
the
Schwinger model with two flavors of equal charge. The
behavior of the
latter is also well known. Again, the expressions
(\ref{corrtSM},
\ref{rise}) are valid with the only difference that now we
have $\mu^2
= 2(g\sqrt{3}/2)^2/\pi$ -- the two flavor loops contribute
in the heavy
 photon mass on the equal footing [and the parameter $g$ in
(\ref{Green}
 -- \ref{rise}) should, of course, be substituted
by $g\sqrt{3}/2$].
 We see that the rising factor (\ref{rise})
now compensates the exponential fall--off of the free
correlator
only
partially, and we have
 \be
 \label{C8}
 C_8(\tau) \propto \exp\{-\pi \tau/L\}
 \ee
 where the subsript 8 indicates the chosen color direction
of the gauge
field background.

 Consider now the case when the gauge field is directed
along the third
color axis. The components $\lambda^{3,8}$ are free and the
components
$\lambda^{1 \pm i2}$, $\lambda^{4 \pm i5}$, and  $\lambda^{6
\pm i7}$
behave in the same way as complex fermions of charge $g$,
$g/2$, and
$g/2$, correspondingly. The problem is reduced to the
Schwinger model
with three flavors of inequal charge. Consider the
correlator
$<\bar \lambda^1 \lambda^1 (\tau) \ \bar \lambda^1 \lambda^1
(0)>$. It has
the same form as before, only the factor $2g^2/\mu^2$
acquires now the value
$$ \frac{2g^2}{\frac{g^2}\pi + \frac{(g/2)^2}\pi +
\frac{(g/2)^2}\pi}\
= \ \frac{4\pi}3$$
  rather than $2\pi$ as in the standard Schwinger model or
$\pi$ as in
the Schwinger model with two flavors of equal charge. We
have
 \be
 \label{C3}
 C_3(\tau) \ =\ \exp\left\{-\frac{2\pi \tau}{3L}\right\}
 \ee
For the components $\lambda^{4 \pm i5},\ \lambda^{6 \pm
i7}$, the
correlator decays faster $\propto \exp\{-5\pi \tau/(3L)\}$,
and their
contribution in the correlator (\ref{corrt}) can be safely
neglected.
   \footnote{The behavior of the correlator on the Euclidean
plane can be
found along the same lines. In the Schwinger model with
several flavors
of arbitrary charges, the  factor (\ref{rise}) rises as a
{\it power}
at large distances. That compensates partially the fall--off
$\sim
x^{-2}$ of the free correlator, and the full correlator for
the fermion
with charge $g_i$ behaves as
$x^{-2\Delta_i}$ with $\Delta_i = 1 - g_i^2/(\pi \mu^2)\  =
\
1 - g_i^2/\sum_i g_i^2$.
The term $g_i^2/(\pi \mu^2)$ is
nothing else as the anomalous dimension of the operator
$\bar \psi_i
\psi_i$. A corresponding conformal theory where this
operator appears
naturally as  a
primary field can be formulated. For example, for two
flavors of equal
charge, it is the primary operator $\cos(\sqrt{2\pi} \chi)$
(or, better to say, a couple of operators $\exp\{\pm
i\sqrt{2\pi} \chi \}$
corresponding to $\bar \psi_{1L} \psi_{1R}$, $\bar \psi_{1R}
\psi_{1L}$
or, equivalently, to $\bar \psi_{2R} \psi_{2L}$ , $\bar
\psi_{2L}
\psi_{2R}$ )
in the
conformal theory of the real massless scalar field $\chi$
\cite{Coleman1}. It is no wonder thereby that one and the
same
factor $\Delta$
determines the power asymptotics of the correlator on the
Euclidean
plane and the exponential asymptotics of the correlator on
the
cylinder. One can map the complex plane on the stripe $0
\leq x \leq
L,\ -\infty < \tau < \infty$ by a conformal transformation
after which
the power behavior of the correlator at large distances is
transformed to
the exponential one (see e.g. \cite{Dotsenko} ) }
Let us compare now the conributions (\ref{C8}) and
(\ref{C3}).
Both decay exponentially at large $\tau$, but
the value of exponent in (\ref{C3}) is smaller than that in
(\ref{C8})
and, at
large $\tau$, the leading asymptotics of the correlator
(\ref{corrt})
is determined by the gauge field background aligned along
the third
color axis and is given by (\ref{C3}).

Notice now that the same result could be obtained from the
hamiltonian
analysis of Ref.\cite{Thies}: Eq.(\ref{C3}) can be
interpreted as
 $$ C(\tau) \ =\ \exp\{-V_\diamond \tau\}$$
 where $V_\diamond$ is the energy (\ref{local}) of the
fourth local minimum
 of the
potential (\ref{Veff3}) discussed before. Indeed, an
accurate treatment
shows that the profile function $f(\tau')$ defined in
(\ref{ab}) for the
saddle point field configuration saturating the path
integral for (\ref{corrt})
rises from 0 to $4\pi/(3L)$ in the region $\tau' \sim 0$
(the width of
this region is of order $g^{-1}$),  stays at this value for
a while until
$\tau'$ approaches the point $\tau$, and goes down to zero
in the region
$\tau' \sim \tau$. But the point
$$ A_1^3 \ =\ \frac{2\pi}{3L}, \ \ \ \ \ \ \ \ A_1^8 = 0 $$
is exactly the point where one of the diamond minima of the
potential
sits. We have for large $\tau$
 \be
 C(\tau) \ =\ |<o|\bar \lambda^a \lambda^a|\diamond>|^2 \
\exp\{-V_\diamond \tau\}
\ee
which coincides with (\ref{C3}).

The instanton (antiinstanton) contribution to the correlator
$C(\tau)$ has the same asymptotic behavior. The relevant
saddle point configuration starts from the central circle in
Fig. 1 at $\tau' = -\infty$. Then at $\tau' \sim 0$ the
field rises in, say, $t^3$ color direction to the diamond
point, stays there for a while, and that provides the
exponent $V_\diamond \tau$ in the asymptotics of the
correlator, after which it does not go back to origin at
$\tau' \sim
\tau$ as in the topologically trivial case, but moves
further
to the closest  triangle or box along the color direction
$(1,0,-1)$ or $(0,-1,1)$ (the  symmetry
axes of the effective potential which are equivalent to
$t^3$ and
correspond to other roots of Lee algebra).

The advantage of the method suggested here is that it can be
easily
generalized for higher $N \geq 4$ where, working in
hamiltonian
approach, we should have studied an intricate
multidimensional
structure of the effective potential
\footnote{We have performed such a study for $N = 4$. The
pattern of
the minima of the effective potential presents an
interesting
3-dimensional lattice akin to the lattice of diamond. But as
it bears
little relevance for the main question studied in this
paper, we will
not distract ourselves here for this issue.}. It turns out
that for any
$N$ the leading asymptotics of the correlator (\ref{corrt})
is due to
the abelian saddle point field configuration (\ref{ab}). In
this background,
off--diagonal components $\lambda^a$ are ``organized'' in a
complex
fermion field $\lambda^{1 \pm i2}$ of the charge $g$ and
$2(N-2)$ complex
fermion fields of the charge $g/2$. The correlator of
$\lambda^{1 \pm
i2}$ components gives the factor
$$ \frac{2g^2}{\pi \mu^2} \ =\ \frac 4N$$
in the exponent and the correlator decays as
 \be
\label{CN}
C_N(\tau) \ \propto \ \exp \left\{- \frac {2(N-2)}N
\frac{\pi \tau}L \right\}
 \ee
Rotating the cylinder where the theory is defined by
$\pi/2$, we arrive
at the conclusion that for $N \geq 3$ at high temperatures
$T \gg g$,
the spatial correlator
 \be
  \label{corrx}
C(x) \ =\ <\bar \lambda^a \lambda^a (x) \ \bar \lambda^a
\lambda^a (0)>_T
 \ee
  decays
exponentially at large distances. By cluster decomposition,
that certainly
implies that
\be
\label{cond0}
<\bar \lambda^a \lambda^a>_{T \gg g,\ N \geq 3} \ \ =\ \ 0
  \ee

\section{Phase Transition.}
\setcounter{equation}0
The bosonization analysis of sect. 4 suggests the presence
of the
fermion condensate in the theory (\ref{Lqcd2}) at $T=0$ for
any $N$ in
the thermodynamic limit ${\cal A} \to \infty$ while the
fermion mass
$m$ is kept small but fixed. On the other hand, as $\ln Z$
does not invovlve a linear term in mass expansion, the
condensate is zero
in the chiral limit $m \to 0$ when the total area of the
manifold
 ${\cal A}$ is kept large but fixed. Also we have seen in
the previous
section that for $N \geq 3$ the condensate is absent at high
temperature $T \gg g$ even in the limit when the length of
the spatial
box $L$ is sent to infinity in the first place. Two options
to resolve
this controversy were mentioned at the end of sect. 4.

One of them postulates  relevance of some non--topological
field
configurations which have only a pair of fermion zero modes
and provide
for a non-zero fermion condensate in the chiral limit. It is
a possible
way out, but it has two obvious weak points. First, we have
no idea on
what these non--topological field configurations are.
Second, assuming
their existence, we do not understand why they disappear at
finite
temperature.

Another option is that the condensate appears at $T = 0$ as
an order
parameter of a spontaneously broken symmetry. In that case,
the limits
{\it i)} ${\cal A} \to \infty$,\ $m$ fixed and {\it ii)} $m
\to 0$,\
${\cal A}$ fixed \ need not commute. The partition function
presents
the sum of  two exponentials (\ref{Zsum}) and the linear
term
$\propto mg{\cal A}$ in the expansion of $Z(m)$ cancels out.

We will argue now that at least for odd $N$, this second
possibility
is rather probable, indeed. First, there {\it is} a
disctrete symmetry
(\ref{Z2Z2}) to be broken. It remains the exact symmetry of
the
lagrangian also on quantum level because instantons involve
a couple of
left--right pairs of zero modes, and the induced 't Hooft
term in the
effective lagrangian $\sim (\lambda_L^a \lambda_R^a)^2$ (we
will first consider the simplest case $N = 3$ )
respects the
symmetry (\ref{Z2Z2}).

Spontaneous breaking of {\it continuous} symmetries is
excluded in 1+1
-- dimensional systems due to Coleman theorem
\cite{Coltheor}, but a
{\it discrete} symmetry can well be broken spontaneously.
The  only
important restriction is that the symmetry should be
restored at any finite
temperature. Really, a physical picture of spontaneously
broken
discrete symmetry involves the presence of the domain walls
between two
different ordered phases. If only one spatial dimension is
there, these
``walls'' present solitons; the corresponding quantum states
have a
finite energy. It is obvious that at finite temperature,
however small
it is, the heat bath involves some number of these
``walls''. And that
exactly means that the vacuum is disordered.

A classical example of a theory involving spontaneous
breaking of $Z_2$
symmetry is one--dimensional Ising model \cite{Ising}.
\footnote{One--dimensional statistical systems correspond to
 1+1 -- dimensional field theories.}
The theory has the hamiltonian
 \be
 \label{Ising}
 H \ =\ -J\ \sum_{i = -N}^{N}\ \sigma_i \sigma_{i+1}\ ,
   \ee
   $N \to \infty$ in the thermodynamic limit. The vacuum
state of
   (\ref{Ising}) is doubly  degenerate:
   $<\sigma> \ = \ 1$ or    $<\sigma>\ = \ -1$. At any non-
zero temperature,
the domain walls (the states with $\sigma_i\ = \ -1$ at $i
\leq n_0$ and
$\sigma_i\ = \ 1$ at $i > n_0$) appear in the heat bath.
Their characteristic
density is $\sim \exp\{-J/T\}$. Thereby, the state is not
ordered
anymore, the correlator $<\sigma_i \ \sigma_{i+M}>$ tends to
zero at $M
\to \infty$, though the spatial correlation length (a
characteristic
value of $M$ when the spin correlator starts to die away) is
exponentially large $\sim \exp\{J/T\}$ when the temperature
is small.
The system has a first--order phase transition at $T = 0$.
\footnote{ Note that
second order phase transitions at $T = 0$ associated with
would--be
spontaneous breaking of a continuous symmetry are also
possible in 1+1
-- dimensional systems. It is exactly what happens in
multiflavor
Schwinger model \cite{Jac}. But as the order parameter is
zero at the
phase transition point and, if $T_c = 0$, there is nothing
below,
the Coleman theorem is not violated.}

Our suggestion is that the same happens in adjoint $QCD_2$
at $N = 3$,
the fermion condensate $<\lambda_L^a \lambda_R^a>$ being the
order
parameter of the symmetry (\ref{Z2Z2}) and playing the role
of
$<\sigma>$.
A number of non-trivial physical consequencies follow from
this assumption.

First, it implies that the  correlation length $l$ of
(\ref{corrx})
rapidly grows as the temperature goes down and becomes
exponentially large
$\sim \exp\{g/T\}$ in the region $T \ll g$. No analytic
calculation in
the region $T \ll g$ is possible. It would be rather
interesting,
however, working still in the region $T \gg g$ where
quasiclassical
approximation applies, to find out what are the {\it
corrections} to the
leading Born--Oppenheimer result [cf. Eq. (\ref{C3})].
 \be
 \label{l}
 l_{T \gg g}  \ =\ \frac 3{2\pi T}
 \ee
 If the first non-leading correction turns out to be
positive, it could
serve as an argument in favor of the scenario that the
corrections
become overwhelmingly large at $T \ll g$.

The second very interesting corollary is that the spectrum
of the
hamiltonian should involve ``walls'' --- the states
interpolating from
the vacuum with negative $<\bar \lambda^a \lambda^a>$ on the
left to
the vacuum with positive $<\bar \lambda^a \lambda^a>$ on the
right. If
the wall states do not exist, but only the states presenting
excitations over the vacuum with $<\bar \lambda^a \lambda^a>
\ \ > 0$ or the
excitations over the vacuum with $<\bar \lambda^a \lambda^a>
\ \ < 0$, we
cannot talk about the spontaneous symmetry breaking in the
physical
meaning of the word. The whole Hilbert state of the system
would be
separated in two subspaces which do not talk to each other,
and a
superselection rule  singling out one of these subspaces
could be
imposed. The situation would be the same as with instantons
in $QCD_4$
\cite{Callan}
or as with adjoint $QCD_2$ at $N = 2$ \cite{cond}. It would
imply the
presence of ``fractons'' like in \cite{fracton} and, as was
mentioned,
it would be difficult to explain where the condensate is
gone at $T
\neq 0$.

Presently, we do {\it not} know whether such wall states
exist. The
spectrum of adjoint $QCD_2$ was studied with some care only
in the
limit $N \to \infty$ \cite{kleb}, but not at finite $N =3$.
\footnote{ However, it is not a
hopeless problem to study the spectrum of the theory on
lattices.
The ``lattice experimental evidences'' in favor or against
our hypothesis
are highly desirable.
Actually, two--dimensional systems are a lot  simpler
than
four--dimensional $QCD$ where  the efforts of lattice people
are mostly
applied. One can only express a wish that the fashion would
change some
time and more lattice works on two--dimensional systems
including
fermions would be done. The field involves many unsolved but
easily
solvable for the experts problems. In the first place, a
number of
exact non-trivial results in the abelian theory (see
\cite{Jac} and
references therein)
should be checked. If theoretical predictions for the
spectrum and
correlators are reproduced in the abelian case, one could
proceed with
two-dimensional non-abelian theories. Also, if numerical
lattice
calculations would reproduce the exact theoretical results
in 2 dimensions,
there would be more trust in
lattice calculations in $QCD_4$ with dynamic fermions.}

The reasoning of this section can be relatively generalized
for higher
odd $N$. The common point is that when $N$ is {\it odd},
the number of zero modes  (\ref{n0}) is always {\it even},
the symmetry
(\ref{Z2Z2}) is not anomalous, and can be broken
spontaneously at
$T=0$. The partition function presents a sum (\ref{Zsum})
of two extensive
exponentials as before. A little bit troublesome point,
however, is
that, say, for $N=5$, the instanton contributions first show
up only in
the quartic term of the expansion of (\ref{Zsum}) in
$mg{\cal A}$. For
$N=137$, they first appear in the term $\sim (mg{\cal
A})^{136}$.
The terms $\sim (mg{\cal A})^2$, \ldots , $\sim (mg{\cal
A})^{134}$ should
come from the path integral in the topologically trivial
sector. Well,
it is somewhat queer, but at least not paradoxical.

The situation with even $N \geq 4$ is more complicated.
The matter
is that in this case the symmetry (\ref{Z2Z2}) {\it is}
anomalous. For
example, for $N = 4$, the field configurations in the
topological class
$k = 1$ involve 3 pairs of zero modes, and the corresponding
't Hooft effective
lagrangian $\sim (\lambda^a_L \lambda^a_R)^3$ is odd under
the transformation
(\ref{Z2Z2}). Generally, the partition function in the
topological sector $k$ acquires the factor $(-1)^k$.
If there is no symmetry, one cannot talk about
its spontaneous breaking. There should be {\it unique}
physical vacuum
state (in a sector with a particular value of discrete
$\theta$ brought about
by instantons) and the equation (\ref{Zsum}) cannot be
written.
Thus, the physics of the theory with odd $N \geq 4$ differs
essentially from
the theory with even $N \geq 4$ (cf. \cite{oddeven}).
In the first case, the hypothesis
about spontaneous $Z_2$ symmetry breaking resolve the
paradox rather
satisfactory (with all reservations given). For large even
$N$, the
paradox is still there, and, at the current level of
understanding,
we  do not dare to speculate  more in this direction.

\section{$O(N^2-1)$ and Disconnected Components.}
\setcounter{equation}0
As far as odd $N$ are concerned, the suggested picture looks
rather self--consistent and nice, and I am
ready to accept bets that $Z_2$ symmetry in the theory with
$N=3,5,\ldots$ is
broken spontaneously, indeed.
There is, however,
a theoretical problem  which is not yet fully understood
and we are in a position to discuss it.

The arguments in favor of the
existence of fermion condensate at $T=0$ come from the
bosonization
analysis. We have interpreted the condensate as the order
parameter of
the spontaneously broken symmetry (\ref{Z2Z2}). The symmetry
(\ref{Z2Z2})
clearly displays itself in the fermion language.
In bozonization language, the corresponding symmetry is
  \be
   \label{hZ2}
   h^{ab} \ \to -h^{ab}
   \ee
   At first sight,
the action (\ref{actnew}) is invariant under the
transformation
(\ref{hZ2}), indeed. The problem is, however, that the
matrix $-h^{ab}$
does not belong to the adjoint representation of $SU(N)$ if
the matrix
$h^{ab}$ does. In particular, the equation
 $$ - \delta^{ab} \ = \ 2\ {\rm Tr} \{u t^a u^{\dagger}
t^b\} $$
 has no solution (it is best seen using the identity ${\rm
Tr}\ h\ = \
|{\rm Tr}\ u|^2 \ - \ 1 \ \geq\ -1$).  Notice now that the
symmetry
(\ref{hZ2}) could be reinforced if assuming $h\ \in \ O(N^2-
1)$ (as Witten
originally suggested for {\it free} fermions). If bosonizing
the theory
with  $h$ belonging to the adjoint representation of the
gauge group
$SU(N)/Z_N$, the transformation (\ref{hZ2}) relates not the
variables
in one and the same bosonized theory, but relates different
theories
corresponding to different subgroups of $O(N^2-1)$. But we
may equally
well multiply $h$ by any matrix of the coset $O(N^2-
1)/[SU(N)/Z_N]$.
All such theories come on the equal footing. We are thus
arriving at
the recipe (\ref{massbR}): the partition function of $QCD_2$
with
massive Majorana fermions is equal to the sum (the integral)
of the
partition functions $Z(R)$ in all possible bosonized
theories
characterized by a matriz $R$.

We cannot {\it prove} now the validity of this recipe.
However, we can
{\it show} that the bosonized partition function with a
particular $R$
has wrong analytic properties as a function of mass. If
summing over
all $R$ with a particular sign prescription (see below), the
correct
analytic properties are reproduced.

Consider first the theory with $N=3$. Let us concentrate on
the
instanton sector and put $R^{ab} = \delta^{ab}$ at first. We
have seen
in Sect. 4 that the leading term in mass expansion of $Z_I$
is $\sim
(mg{\cal A})^2$. Consider now the next term $\propto m^3$.
It appears
 when pulling down the mass term in the action thrice.
Proceeding along the same lines as in Sect. 4 (i.e. taking
into account
only the zero Fourier harmonic $u_0$ and imposing the
requirement
$[u_0,\ T^*] = 0$ ), we obtain
  \be
  \label{Zm3}
  Z_I^{N=3} \ =\ C_2 (mg{\cal A})^2 \ + \ C_3 (mg{\cal A})^3
   \int du^{(2)}  |{\rm Tr}\ u^{(2)}|^2
   \left({\rm Tr}\ u^{(2)}\right)^2\
+ \ O(m^4)
\ee
The group integral in  (\ref{Zm3}) is nonzero and we get a
nonzero
cubic term in the expansion of $Z_I$ in mass.

However,  the cubic term is absent in the original fermion
theory.
Really, mass dependence comes from the fermion determinant
  \be
    \label{Det}
    Det_{Majorana}^{N=3} ||i\not\!\!{\cal D} + m||\  =   \
\left[
     Det_{Dirac}^{N=3} ||i\not\!\!{\cal D} +
m||\right]^{1/2} \ \sim
     m^2 \prod_n'(m^2 + \lambda_n^2)
     \ee
where the product runs over all nonzero eigenvalues of the
Euclidean
Dirac operator, only one eigenvalue of each doubly
degenerate pair
being taken into account \cite{cond}. The determinant
(\ref{Det})
involves only even powers of $m$.

It is easy to see that, if allowing for an arbitrary $R \in
O(8)/[SU(3)/Z_3]$ and integrating over $R$, the expression
 \be
 \label{Ztrue}
 Z_I^{true} \ =\ \int dR\ Z_I(R)
 \ee
 also involves    only even powers. For each $R$, the theory
with $R' =
-R$ also contributes in the integral. But the mass terms
(\ref{massbR})
in these two theories have opposite sign.

Consider now a theory with even $N$. The case $N=2$ is
already non-trivial. The symmetry (\ref{hZ2}) is realized on
the full $O(3)$ group involving two disconnected components
$SO(3)$ where the bosonized theory (\ref{actnew}) is
formulated. We have to take into account the contributions
of both components in the partition function. But, in
contrast to $N=3$, it would be incorrect just to sum up the
corresponding contributions. Speaking precisely, it is
correct in the topological trivial sector, but not in the
instanton sector.

The contribution of the component with $Det\ ||h|| \ = \ 1$
in the partition function in the instanton sector is
  \be
  \label{ZIpl}
Z_I^{N=2}(+)  = C_1 \ mg{\cal A} + C_2 \ (mg{\cal A})^2 \ +
\ O(m^3)
 \ee
with a nonzero $C_2$ given by the integral
$$C_2 \ \propto \ \int_0^{2\pi} d\beta_3 \exp\{-i\beta_3\}
(2\cos \beta_3 + 1)^2 \ \neq \ 0$$
Like in the previous case, it has wrong analytic properties
involving both odd and even powers of mass. The mass
dependence of $Z_I$ in the fermion theory comes from the
Majorana fermion determinant which involves for $N=2$ only
odd powers:
$$  Det_{Majorana}^{N=2} ||i\not\!\!{\cal D} + m|| \ \sim m
\prod_n'(m^2 + \lambda_n^2) .$$
To reproduce this behavior, we have to {\it subtract} the
contribution $Z_I^{N=2}(-)$ of the odd $SO(3)$ component
with
   $Det\ ||h|| \ = \ -1$ from (\ref{ZIpl}). The
corresponding
theory differs from the theory of the even $SO(3)$ component
only by the sign of the mass term (\ref{massbR}).
The expansion of  $Z_I^{N=2}(-)$ in mass has exactly the
same form as (\ref{ZIpl}) up to the opposite sign of odd
powers. We are defining now
 \be
 Z_I^{N=2}(true) \ = \  Z_I^{N=2}(+)\ -\ Z_I^{N=2}(-)
 \ee
$Z_I^{N=2}(true)$ involves only odd powers of mass. Our
hypothesis is that it exactly corresponds to the instanton
partition function of the fermion theory.

Consider now a general case. Let first $N$ be odd. The
number of zero mode pairs $k(N-k)$ is even for any $k$ and
the expansion of the partition function in mass in the
topological sector $k$ starts with $m^{k(N-k)}$ and involves
only even powers of $m$. The expansion of the partition
function $Z_k$ in the bosonized theory (\ref{actnew}) with
the mass term (\ref{massb}) also starts with  $m^{k(N-k)}$
[see Eq. (\ref{ZIk})], but includes both even and odd
powers. For odd $N$, the group $O(N^2-1)$ includes only one
connected component.
The same arguments as for the case $N=3$ considered before
show that the odd powers of mass cancel out in the integral
(\ref{Ztrue}) over the theories with different $R$. This
integral should correspond to the partition function $Z_k$
in the original fermion theory.

Let now $N$ be even. The value $k(N-k)$ may be odd or even
depending on $k$. For example, for $N=4$, the sectors
$k=1,3$ involve 3 pairs of fermion zero modes, and the
sector $k=2$ involves  4 such pairs. In the former case, the
expansion of $Z_k^{ferm}$ involves only odd powers of mass,
and in the latter case --- only even powers. On the other
hand, the mass expansion of $Z_k$ in the bosonized theory
with the mass term (\ref{massb}) includes both even and odd
powers for any $k$. Note now that the group $O(N^2-1)$
includes 2 disconnected components for even $N$. Our recipe
 reads
   \be
  \label{recipe}
 Z_k^{N\ even}(true) \ = \ \int \ dR_+ Z_k^{N \ even}(R_+)\
+\ (-1)^k\ \int\ dR_- \ Z_k^{N\ even}(R_-)
 \ee
 The odd (even) powers of mass cancel out in the integrated
partition function (\ref{recipe}) with even (odd) $k$ and
the correct analytic properties of $Z_k$ are reproduced.

 Again, we see the distinction between odd and even $N$.
Obviously, there is a relation between the existence of two
disconnected components in  $O(N^2-1)$ for even $N$ and the
fact that the symmetry (\ref{Z2Z2}) is anomalous. Indeed,
the partition function (\ref{recipe}) is  invariant over the
bosonic counterpart of this symmetry, the transformation $h
\to -h$, for even
$k$ but not for odd $k$.

\section{Discussion.}
\setcounter{equation}0
The main physical signature of the suggested scenario with
spontaneous breaking of discrete $Z_2$ symmetry is the
presence of the domain wall solitons --- the states which
interpolate between different vacua --- in the spectrum of
the theory. If the domain walls are absent, different vacua
are completely unrelated to each other and belong to the
different sectors of Hilbert space. In that case, a
superselection rule which selects a particular sector once
and forever in the whole physical space should be imposed.
Then there is no  spontaneous symmetry breaking in the
physical meaning of this word. This is the situation in
standard $QCD_4$ (the vacuum involves a continuous
degeneracy in $\theta$, but one cannot talk of the
spontaneous breaking of $U(1)$ symmetry because the physical
signature of this breaking --- the massless Goldstone boson
which is singlet in flavor --- is absent). This is also a
situation in pure YM theory at high temperature where the
physical domain walls interpolating between different $Z_N$
``phases'' are absent and one cannot talk about spontaneous
breaking of $Z_N$ discrete symmetry \cite{bub}. And this is
the situation in adjoint $QCD_2$ with $N=2$ where two
sectors (\ref{Zpm}) are not physically related and there are
no walls.

The fact that we cannot at present establish the existence
of  domain walls in adjoint $QCD_2$ with $N \geq 3$
explicitly is the main reason why we are still talking about
the possibility of spontaneous breaking of $Z_2$ symmetry in
this theory (even for odd $N$ where the symmetry
(\ref{Z2Z2}) to be broken is retained on the quantum level)
without metal in voice.

Two--dimensional model considered in this paper presents an
interest on its own, but the main point of interest are the
lessons one can learn from the analysis of this model for
4--dimensional supersymmetric Yang--Mills theories. These
theories attracted recently a considerable attention after
appearance of the paper of Witten and Seiberg who calculated
exactly the spectrum of physical states in ${\cal N} = 2$
supersymmetric Yang--Mills theory \cite{Seiberg}.

There is a long--standing unresolved problem in a more
simple ${\cal N}=1$ supersymmetric Yang--Mills theory
involving only gluons and gluinos. Supersymmetric Ward
identities display the constant (x-independent) behavior of
the fermion correlator
  \be
  \label{corr4N}
< \lambda^a_\alpha \lambda^{a\alpha}(x_1)\ \ldots \
 \lambda^a_\alpha \lambda^{a\alpha}(x_{N})>\ = \ {\rm const}
  \ee
(for $SU(N)$ gauge group).
Instanton calculations (which are valid at small $|x_i -
x_j|$) show that that constant is nonzero \cite{Shif}. That
implies the presence of gluino condensate. However, standard
instantons involve $2N$ fermion zero modes and, assuming
that only instantons contribute  and the extensive form
(\ref{extens}) of the physical partition function with only
one physical vacuum state is valid, we are led to the same
contradiction as in adjoint $QCD_2$ at $N \geq 3$ considered
in this paper --- the linear in mass term in Taylor
expansion of  partition function, which should be there due
to the presence of non-zero linear  term in Taylor expansion
of $\epsilon_{vac}(m)\ \equiv$ the fermion condensate,
cannot
be reproduced.

Just as in adjoint $QCD_2$, there are only two ways out.
Either {\it i)} we should assume that $Z_{2N}$ symmetry in
SYM lagrangian (a remnant of $U(1)$ symmetry after taking
anomaly into account) is broken spontaneously
 down to $Z_2$ or {\it ii)} that an additional
superselection rule should be imposed. It amounts to
allowing the $\theta$ parameter to vary within the interval
 \be
  \label{thetaN}
\theta \ \in \ (0,\ 2\pi N)
  \ee
 In the first case, the physical domain walls separating
different $Z_N$ phases should be present in the theory. In
the second case, the ``phases'' should be completely
unrelated and the domain walls must be absent.

As far as SYM theory with $SU(N)$ gauge group is concerned,
we favor more the second possibility. After all, at least in
toroidal geometry, the Euclidean configurations with
fractional topological charge $\propto 1/N$ appear on an
equal footing with instantons \cite{toron} and an additional
superselection rule with respect to a large gauge
transformation changing the Chern--Simons number by $1/N$
arises quite naturally. Actually, one can explicitly
calculate the toron contribution in the partition function
of the theory at finite volume \cite{Leut}. There are also
additional arguments coming from the analysis of the pure
Yang--Mills theory  in large $N$ limit. If no fermions are
there, the partition function is a non-trivial function of
$\theta$. At large $N$, a smooth $\theta$ -- dependence of
the partition function can be achieved only if allowing
$\theta$ to vary within the interval (\ref{thetaN})
\cite{Leut}. All together that make us to believe that the
superselection rule leading to the classification
(\ref{thetaN}) should be imposed,  there are no walls and no
spontaneous symmetry breaking.

For a proper balance, we should also mention
counterarguments to this scenario.
\begin{enumerate}
\item Toron configurations can be written in a finite
toroidal box but not in $S^4$ or $S^3 \times R$ geometry.
 If we do not restrict ourselves with fiber bundles on
compact manifolds, meron solutions
with fractional topological charge which live in $R^4$ and
have a singular field strength at one point can
be written \cite{meron}. They have an infinite action, but
still may be relevant for physics \cite{Gross}. Torons on
tori are not similar to merons in flat space and to the
absence of anything on a sphere.
 The physics, however, should not depend on boundary
conditions if the box is large enough.
\item In contrast to instantons, toron configurations are
delocalized. Again, we cannot visualize at present how these
delocalized configurations manage to contribute in local
physical quantities.
\footnote{A counterargument to this counterargument can also
be suggested. Really, {\it classical} instanton solutions in
Schwinger model are also delocalized, but still instantons
contribute to local observables like the fermion condensate
\cite{indus,Wipf,inst}. Anyway, we understand the mechanism
of that in the Schwinger model --- after taking into account
the fermion determinant, a relevant saddle point of the
corresponding path integral presents a localized vortex-like
configuration \cite{inst} [cf. Eq. (\ref{A0k}) and the
discussion thereafter]. But we do not understand it in
the 4--dimensional SYM theory which we would like to.}
\item An argument in favor of existence of the walls in
$SU(N)$ theory can be put forward if considering the ${\cal
N} = 1$ theory with matter fields (supersymmetric $QCD$).
When the mass of quarks and squarks is small, the theory is
in weak coupling Higgs phase (see e.g. \cite{UFN}). The
different $Z_N$ phases are associated with different values
of the Higgs average and the domain wall solitons with
finite energy density interpolating between different Higgs
phases probably exist. One can send then the mass of matter
fields to infinity after which they decouple. A
renormalization group analysis seem to show that the energy
density of these walls remains finite also in this limit
which means the existence of physical walls also in pure SYM
theory \cite{Shifwall}.
\end{enumerate}

As I already mentioned, my own  guess is that the arguments
{\it pro} overweigh in this case the arguments {\it contra}
and the walls are not really there in $SU(N)$ theory. But
this
guess does not have the rank of a  statement. Obviously,
more study of the question is necessary.

The situation is, however, different in the theories with
higher orthogonal and exceptional gauge groups. Again,
supersymmetric Ward identities and instanton calculations
imply that the $d$ - point function of several fermion
scalar densities like (\ref{corr4N}) ( $d$ is the Dynkin
index of the group. For higher
orthogonal groups $SO(N \geq 5)$, $d \ =\  N-2$ .) is a non-
zero constant \cite{SYMort}.  That
implies the presence of the fermion condensate, but, in
contrast to theories with unitary groups, no toron
configurations with fractional topological charge which
could generate the condensate explicitly are known. In that
case, the option involving spontaneous breaking of $Z_d$ -
symmetry looks much more probable. The domain walls should
exist.

We think that the further study of adjoint $QCD_2$ for $N
\geq 3$ would make a lot of sense. This 2D theory  is much
simpler than 4D SYM theories. One can hope that a
definite answer to the question whether domain walls exist
in two dimensions (we believe they do) would be obtained
reasonably soon. The resolution of this question could
provide crucial insights on what happens in four dimensions.

 \section*{Acknowledgments.}
I am deeply indebted to Nikita Nekrasov who has shown me the
form
(\ref{actnew}) for the gauged WZNW action
and to Michael Shifman for extensive and fruitful
discussions. This work has been done under the partial
support of INTAS
Grants CRNS--CT 93--0023, 92--283, and 94--2851.

\end{document}